\documentclass[a4paper,twocolumn,11pt,submitted=2025-07-27]{quantumarticle}
\pdfoutput=1
\usepackage[utf8]{inputenc}
\usepackage[english]{babel}
\usepackage[T1]{fontenc}
\usepackage{amsmath}
\usepackage{hyperref}
\usepackage{tikz}
\usepackage{lipsum}
\usepackage{float}
\usepackage[utf8]{inputenc}

\usepackage{orcidlink}

\usepackage[T1]{fontenc}
\usepackage{amsmath}
\usepackage{hyperref}
\usepackage{physics}

\usepackage{longtable}
\usepackage{array}
\usepackage[margin=1in]{geometry}
\usepackage{ragged2e}

\begin{document}

\title{Pushing the Limits of LLMs in Quantum Operations}

\newcommand{\orcidauthorA}{0000-0003-4555-777X} 
\newcommand{\orcidauthorB}{0000-0002-3502-7081} 

\author{Dayton C. Closser\,
  \href{https://orcid.org/0000-0003-4555-777X}{%
    \includegraphics[height=1.5ex]{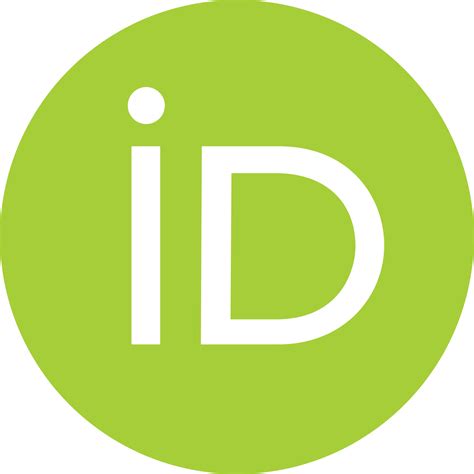}%
  }%
}
\affiliation{dayton.closser@duke.edu}
\affiliation{Department of Electrical \& Computer Engineering, Pratt School of Engineering, Duke University, Durham, North Carolina, 27708, United States of America}
\orcid{0000-0003-4555-777X}
\author{Zbigniew J. Kabala\,
  \href{https://orcid.org/0000-0002-3502-7081}{%
    \includegraphics[height=1.5ex]{logo-orcid.jpg}%
  }%
}
\affiliation{zbigniew.kabala@duke.edu}
\affiliation{Department of Civil \& Environmental Engineering,  Pratt School of Engineering, Duke University, Durham, North Carolina, 27708, United States of America}
\orcid{0000-0002-3502-7081}
\maketitle

\begin{abstract} 
{What is the fastest Artificial Intelligence Large Language Model (AI LLM) for generating quantum operations? To answer this, we present the first benchmarking study comparing popular and publicly available AI models tasked with creating quantum gate designs. The Wolfram Mathematica framework was used to interface with the 4 AI LLMs, including WolframLLM, OpenAI ChatGPT, Google Gemini, and DeepSeek. This comparison evaluates both the time taken by each AI LLM platform to generate quantum operations (including networking times), as well as the execution time of these operations in Python, within Jupyter Notebook. Our results show that overall, Gemini is the fastest AI LLM in producing quantum gate designs. At the same time, the AI LLMs tested achieved working quantum operations 80\% of the time. These findings highlight a promising horizon where publicly available Large Language Models can become fast collaborators with quantum computers, enabling rapid quantum gate synthesis and paving the way for greater interoperability between two remarkable and cutting-edge technologies.}
\end{abstract}
\section{Introduction}

Today's quantum computers are reaching new milestones in performance and efficacy. These systems, known as Noisy Intermediate Scale Quantum (NISQ) computers, vary in size "with a number of qubits ranging from 50 to a few hundred \cite{Preskill2018quantumcomputingin}." Key challenges confronting NISQ systems include reliability, scalability, and, soon we posit, interoperability. 

Reliability-wise NISQ systems face complex quantum noise environments. The term “Noisy emphasizes...imperfect control over those qubits; the noise will place serious limitations on what quantum devices can achieve in the near term \cite{Preskill2018quantumcomputingin}." When noise undesirably meshes with quantum information, that information becomes "entangled" with the environment in a process known as decoherence. Approaches to counter undesirable entanglement with the environment, or decoherence induced errors in quantum computers is addressed by the field of quantum error correction. This is a process whereby quantum information is encoded into a larger subspace of Hilbert space by monitoring symmetries of code space \cite{brown2023}. The end-goal is to achieve "fault-tolerance...the property that a circuit overall can be more reliable than the faulty gates that make it up \cite{McEwen2023relaxinghardware}." This is especially relevant in "large-scale quantum computing [which] requires quantum error correction and fault-tolerance protocols to control the growth of errors \cite{fahimniya2023fault}."
\begin{quote}
\textit{"The only qubit that has no errors is the qubit which never does anything."} \\
--- Kenneth R. Brown, \textit{Quantum Error Correction and Architectures \cite{brown2024quantum}}
\end{quote}
Scalability-wise, the hardware of NISQ systems face challenges with control and confinement or again, coupling to the environment. This warrants approaches as atomic, molecular, and optical physics and human-crafted hardware \cite{brown2023}. 
It should be mentioned here that some entanglements are desirable by design;
"the power of quantum parallelism [multiple operations] relies on the phenomenon \cite{mcintyre2022quantum}."
Entanglement plays an "intriguing role...in quantum computing. As with most good things, it is best consumed in moderation \cite{PhysRevLett.102.190501}."

While much research in the field focuses on the former two factors, we propose a third critical factor for quantum computers: interoperability, or the ability of quantum systems to effectively collaborate and communicate with other technologies and users. This becomes vital when considering who provides the information quantum computers are given, that is, the quantum operations and gate designs issued. Historically, the design of quantum circuits has relied heavily on human expertise. However, the advent of accessible Artificial Intelligence Large Language Models (AI LLM, or LLM for short) and machine learning methodologies presents a compelling opportunity to systematically explore and generate quantum circuit architectures that implement desired quantum operations. We argue this third factor of interoperability should be seriously considered, especially "given the successes of both machine learning and quantum computing, combining these two strands of research is an obvious direction \cite{arunachalam2017surveyquantumlearningtheory}." 

Although research at the intersection of LLMs and quantum computing is still emerging, a crucial gap remains: a rigorous evaluation of publicly available AI models for quantum design applications. Our paper addresses this gap by delivering one of the first systematic benchmarks of leading AI models such as WolframLLM, OpenAI, Gemini, and DeepSeek, focused specifically on quantum circuit design and gate synthesis. In summary, such systematic comparisons of AI models specifically benchmarking fast quantum design and gate synthesis have been limited and nascent, until now. 

\section{Literature Review}

The Web of Science is the gold-standard database of peer-reviewed journal articles. At title search within it for "quantum comput$*$" (where “$*$” stands for a wild card, comput covers computation, computational, computer, computers, etc.) yields 8,139 entries with the earliest published in 1975. Even though still relatively young, the field of quantum computing or, more generally, quantum information science, is maturing rapidly with a number of research and review articles published in top journals.

In Nature, Ladd et al. (2010) explored whether storing, transmitting, and processing information encoded in uniquely quantum systems is feasible, noting that while promising, it remains unclear which technology will prevail \cite{Ladd2010}. In Science, O’Brien (2007) highlighted that all-optical quantum computing became feasible with single-photon sources and detectors as early as 2001, though practical scaling presents challenges \cite{OBrien2007}. Belenchia, Wald, and Giacomini argued "that a quantum massive particle should be thought of as being entangled with its own Newtonian-like gravitational field, and thus that a Newtonian-like gravitational field can transmit quantum information \cite{belenchia2019information}."

Wasielewski et al. (2020) in Nature Reviews Chemistry emphasize molecules' quantum properties as new avenues for advancing quantum information science and its applications \cite{Wasielewski2020}.  Bahrami, et al. "computed the decoherence rate for two quite common types of environment: thermal radiation and background gas \cite{bahrami2012decoherence}. Additionally, Tóth and Apellaniz discussed applications for quantum Fisher information, such as "how it can be used to obtain a criterion for a quantum state to be a macroscopic superposition \cite{Toth2014}." Nayak et al. (2008), in Reviews of Modern Physics, focus on topological quantum computation using non-Abelian anyons, which offers fault-tolerance through nonlocal encoding \cite{Nayak2008}. 

Trapped-ion quantum computing, reviewed in Applied Physics Reviews by Bruzewicz et al. (2019), shows promise with few-ion systems already demonstrating quantum algorithms \cite{Bruzewicz2019}. Reiher et al. (2017) argued in PNAS that quantum devices are nearing computational power beyond classical supercomputers and are well suited for studying complex chemical reactions \cite{Reiher2017}. 

Raussendorf and Briegel (2001) proposed one-way quantum computers based on cluster states, where computation occurs via one-qubit measurements \cite{Raussendorf2001}. And famously, John Preskill (2018) coined the name of Noisy Intermediate-Scale Quantum (NISQ) devices, and in his paper, described the upcoming era of 50-100 qubit machines to outpace classical computers, though still fully expecting fault-tolerant computing to require improved gates first to accomplish this feat \cite{Preskill2018quantumcomputingin}. 

Bourassa et al. (2021) advocated photonics as a modular, room-temperature platform for scalable, fault-tolerant quantum computing \cite{Bourassa2021}, while Steiger et al. (2018) introduced ProjectQ, an open-source framework to develop and simulate quantum algorithms \cite{Steiger2018}. 

Lohani et al. (2022) proposed data-centric machine learning methods that enhance quantum state reconstruction accuracy without changing model architectures \cite{Lohani2022}. Tóth and Apellaniz (2014) also reviewed quantum metrology advances and how noise limits precision \cite{Toth2014}. Loss and DiVincenzo (1998) discussed universal quantum gates using spin states in coupled quantum dots \cite{Loss1998}. Kitaev (2003) described fault-tolerant 2D quantum computation using anyonic excitations and their braiding operations \cite{Kitaev2003}. 

In Current Opinion in Structural Biology, Lappala (2024) highlighted how integrating AI, machine learning, and quantum computing is revolutionizing molecular dynamics simulations, calling for multidisciplinary approaches to meet emerging challenges \cite{LAPPALA2024102919}. 

And in their works, senior research scientists Srinivasan Arunachalam and Ronald de Wolf hinted at interoperability in several ways: This included using "quantum machine learning for “quantum supremacy”, i.e., for solving some task using 50–100 qubits in a way that is convincingly faster than possible on large classical computers \cite{arunachalam2017surveyquantumlearningtheory}." In essence, could certain "practical machine learning problems [dovetail] with a large provable quantum speed-up? \cite{arunachalam2017surveyquantumlearningtheory}." 

At the same time, understanding the potential impacts of interoperability is also imperative, especially considering "the integration of artificial intelligence and quantum computing...could introduce new issues related to data quality \cite{LAPPALA2024102919}."

Other researchers have also suggested interoperability between LLMs and quantum computers: work by Nicolas Dupuis et al. has discussed "an area of interest...to develop specialized LLMs for quantum code generation \cite{dupuis2024qiskitcodeassistanttraining}."

Furthermore, the authors recognized, "there is a noticeable gap in the application of machine learning and classical intelligence systems and algorithms to augment quantum ecosystems and platforms and empowering quantum computing practitioners \cite{dupuis2024qiskitcodeassistanttraining}." 

As well, Sanjay Vishwakarma et al. employed Qiskit HumanEval for "quantum computing tasks, each accompanied by a prompt, a canonical solution, a comprehensive test case, and a difficulty scale to evaluate the correctness of the generated solutions \cite{10821459}." Lastly, Basit et al. introduced "a novel, high-quality dataset comprising 3,347 PennyLane-specific quantum code samples and contextual descriptions, specifically curated to support LLM training and fine-tuning for quantum code assistance \cite{basit2025pennylangpioneeringllmbasedquantum}."
\section{Materials and Methods}
The experiment's first step involved creating a master quantum prompt table (available individually in the Appendix as Table \ref{tab:quantum_part1}, Table \ref{tab:quantum_part2}, Table \ref{tab:quantum_part3}, Table \ref{tab:quantum_part4},
Table \ref{tab:quantum_part5}), which lists every experiment that would be input. The concepts chosen were an array of topics pulling from quantum gates (such as the Pauli gate set), to experiments in the quantum computing field (such as quantum fourier transforms). Twenty-five prompts were elected and chosen to get a sense of how LLMs would handle the subject materials. This number was primarily chosen because of the limitations to how much execution time is permitted on each LLM service respective to cost and the project's budget.

The Wolfram Mathematica framework was used to interface with the 4 LLM AIs. Their programmatic results were saved to text files, then carefully preserved, copied, then filtered into text files that could be executed in the Python3 language. This entailed painstakingly and carefully removing content such as header information that was not directly called for as part of the programmatic prompt. The removal was careful to only carve out content specifically between the comments START and END, as requested for in the prompts. The programmatic results were then executed in a Jupyter Notebook environment. The test platform computer was a Red Hat Enterprise Linux laptop running an Intel i7-7700HQ (3.80 GHz) with 16 gigabytes of ram.

\begin{figure}[H]
    \centering
    \includegraphics[width=1.0\linewidth]{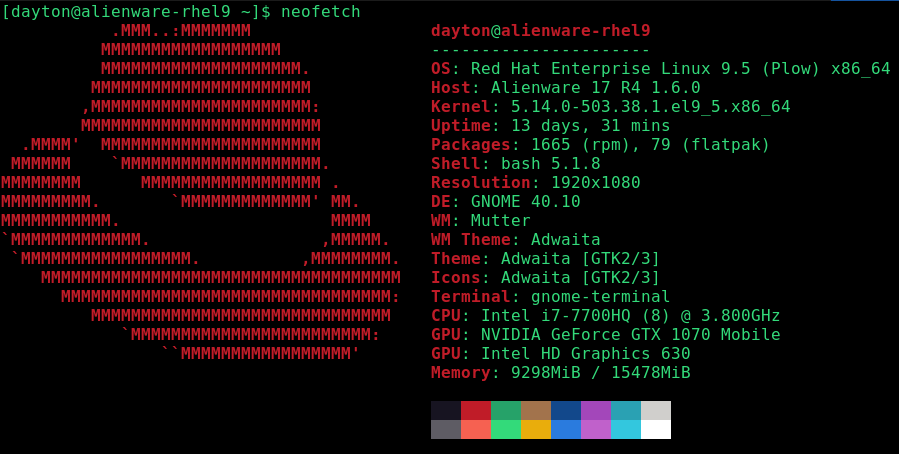}
    \caption{Testbed Platform.}
    \label{fig:OS}
\end{figure}

As well, we also measured the quantities of time for how long each AI platform took to respond to initial requests or connections with Mathematica. At the time of this writing, this was not possible for Wolfram LLM, as there was no native API support for measuring the connection time for this internal Wolfram service. Crucially, this captured the total time from the sending of the prompt request to the LLM, the "thinking time of the LLM" and the response to Mathematica. We also solely measured the connection time to the LLMs themselves, to get a gist of the actual LLM "thinking" time, versus networking latencies.

Then, we compared each of the LLM's programmatic output by the execution times of prompts on the testbed machine with a test harness we devised to benchmark the performance of the output. We evaluated the overall execution time in seconds. Furthermore, we looked at success rates of each prompt measured by the output created and whether or not it would compile with our test harness and compiler. If not, which ones did not compile? Lastly, we investigated the output to see how it evaluated to the expected criteria. The following results are from Wolfram Mathematica, Jupyter Notebook, Python3, and GenAI to produce open source graphics from the corresponding data.
\section{Results}
We first graphed the results of all the LLM models, with an optimal window between 0-15 seconds. 3 of the 4 LLM models generated consistently and relatively similarly low execution times for synthesizing quantum gates (Gemini, WolframLLM, and OpenAI. In contrast, DeepSeek was the relative slowest with execution times that exceeded the 15 second window. We ranked the order of the LLMs here by on average (number of peaks) fastest to slowest. Important to note, this also included the time to file the request with the LLM service, as well as the "return trip" of the prompt.

\begin{figure}[H]
    \centering
    \includegraphics[width=1.0\linewidth]{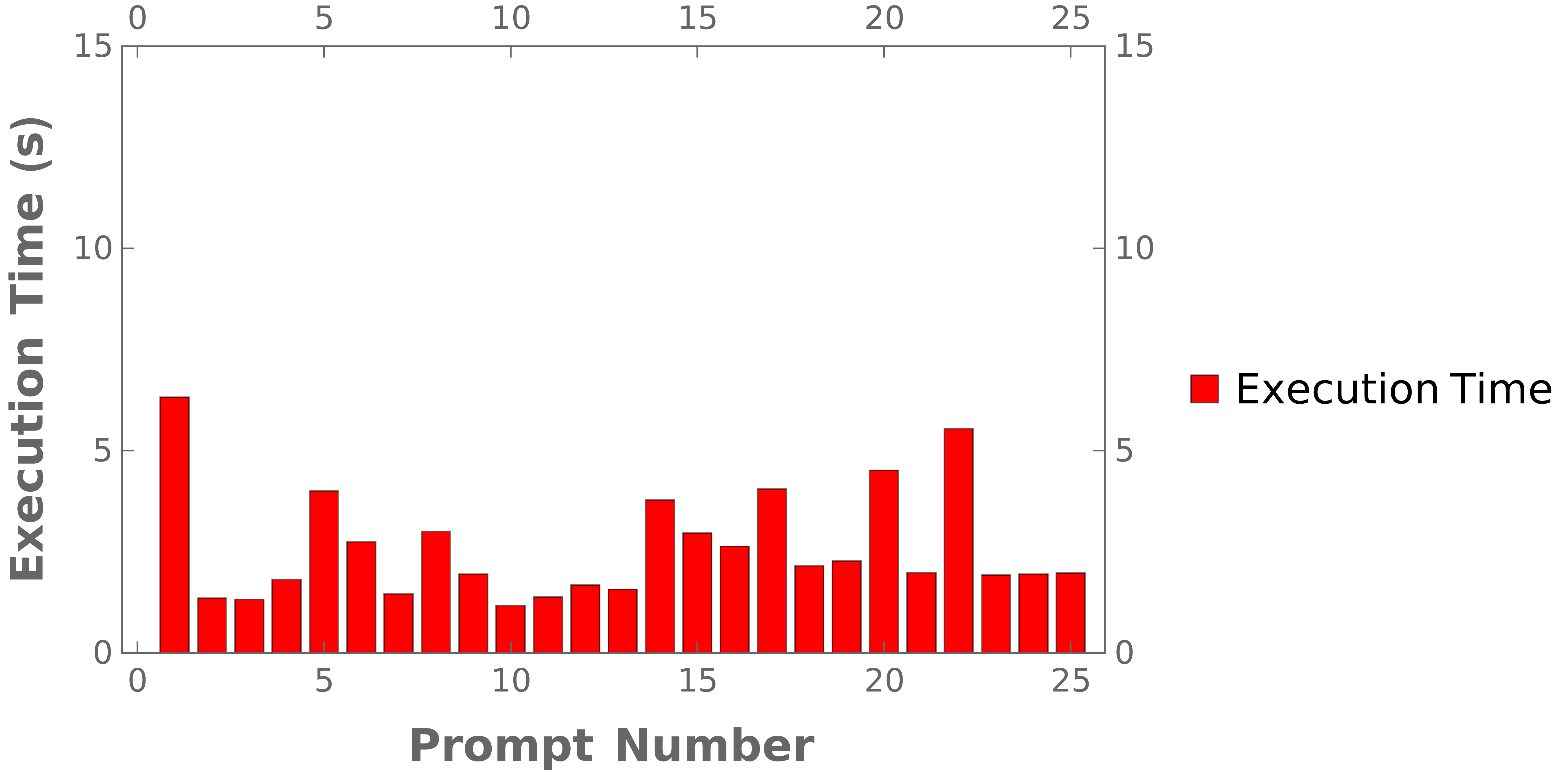}
    \caption{Gemini AI Execution Times.}
    \label{fig:OS}
\end{figure}

Gemini was the fastest LLM on average to execute our prompts.

\begin{figure}[H]
    \centering
    \includegraphics[width=1.0\linewidth]{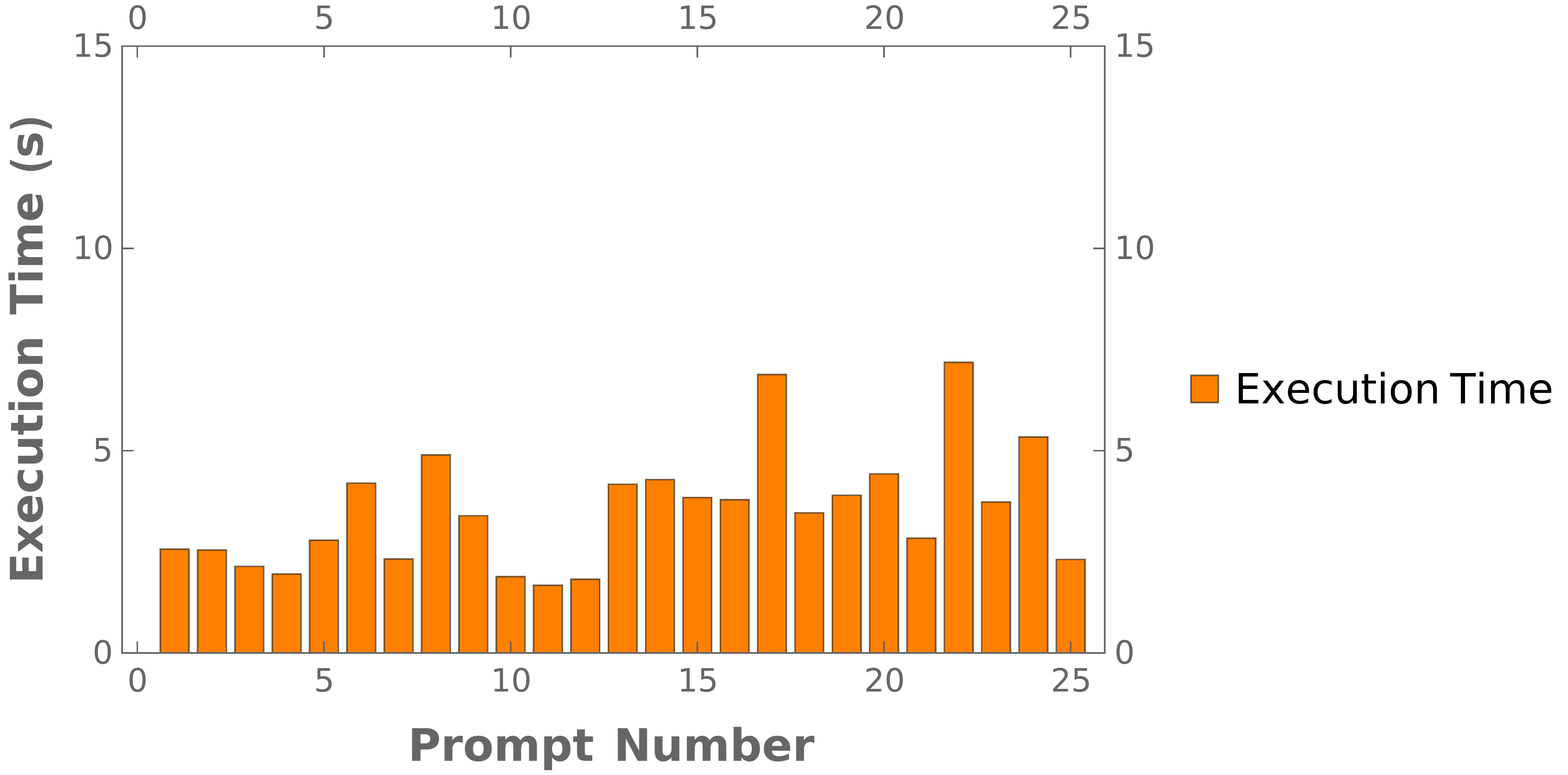}
    \caption{WolframLLM AI Execution Times.}
    \label{fig:OS}
\end{figure}

\begin{figure}[H]
    \centering
    \includegraphics[width=1.0\linewidth]{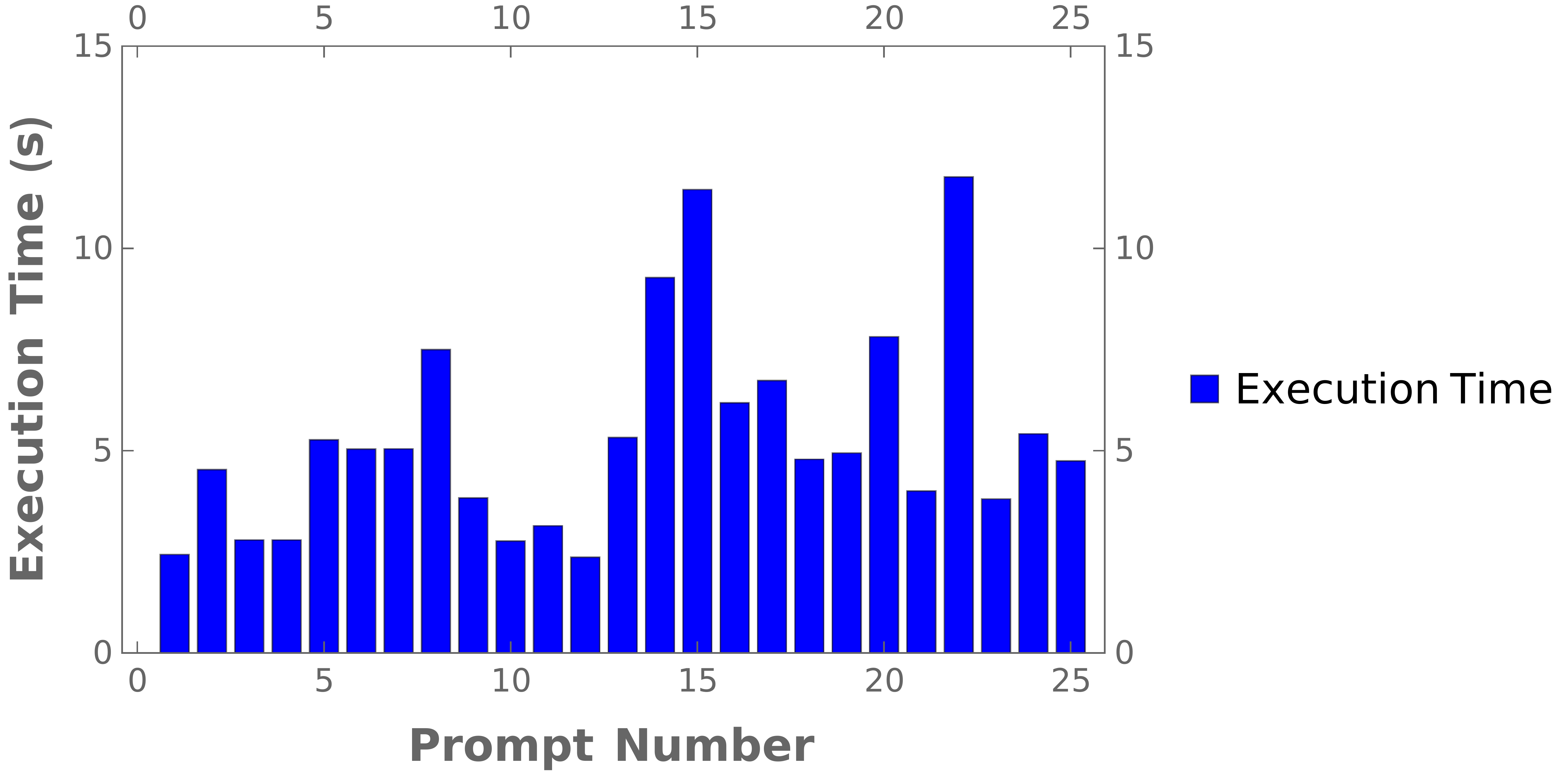}
    \caption{OpenAI Execution Times.}
    \label{fig:OS}
\end{figure}

\begin{figure}[H]
    \centering
\includegraphics[width=1.0\linewidth]{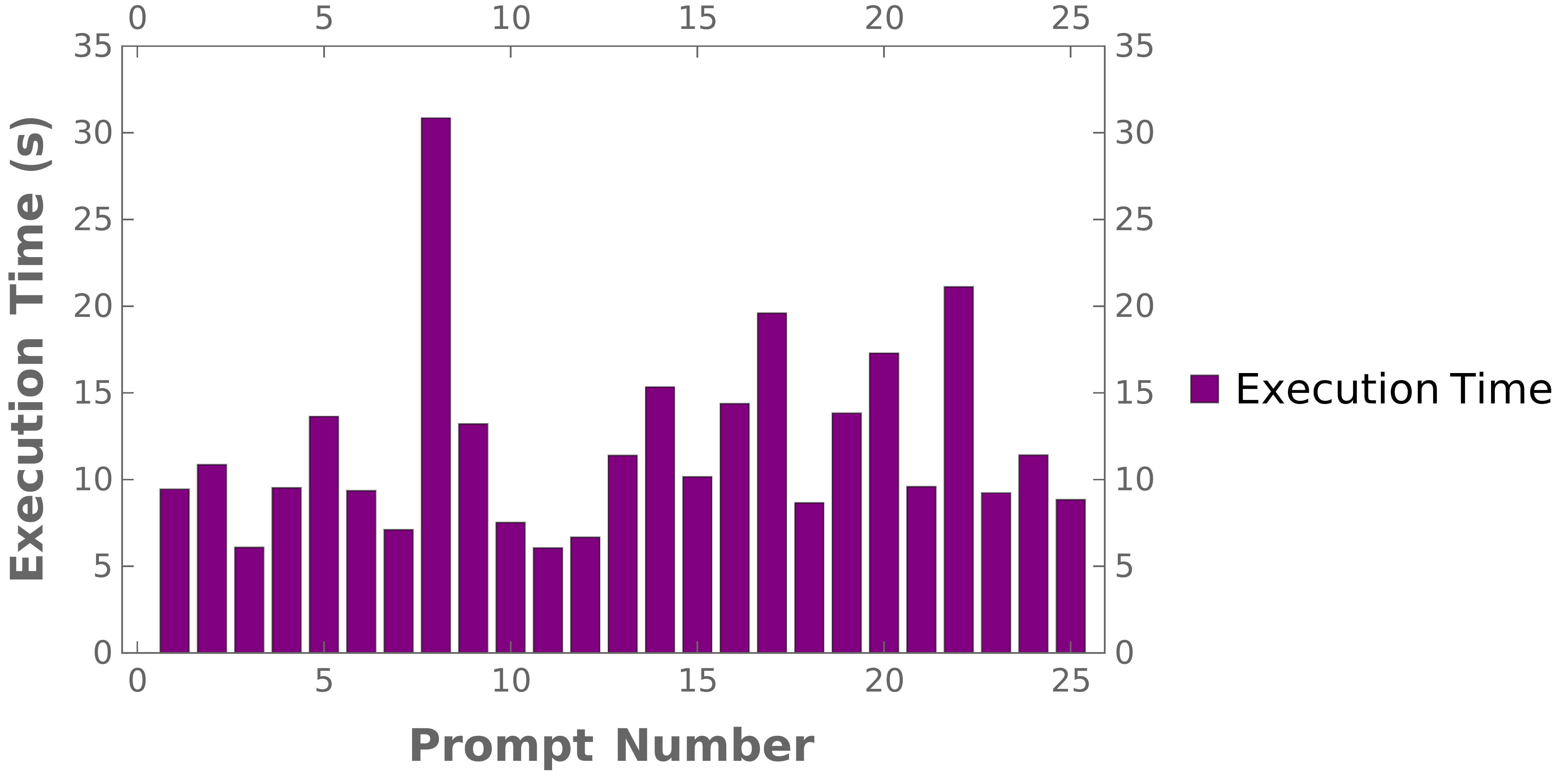}
    \caption{DeepSeek AI Execution Times.}    \label{fig:DeepSeekAIExecutionTime}
\end{figure}

\begin{figure}[H]
  \centering
  \includegraphics[width=0.45\textwidth]{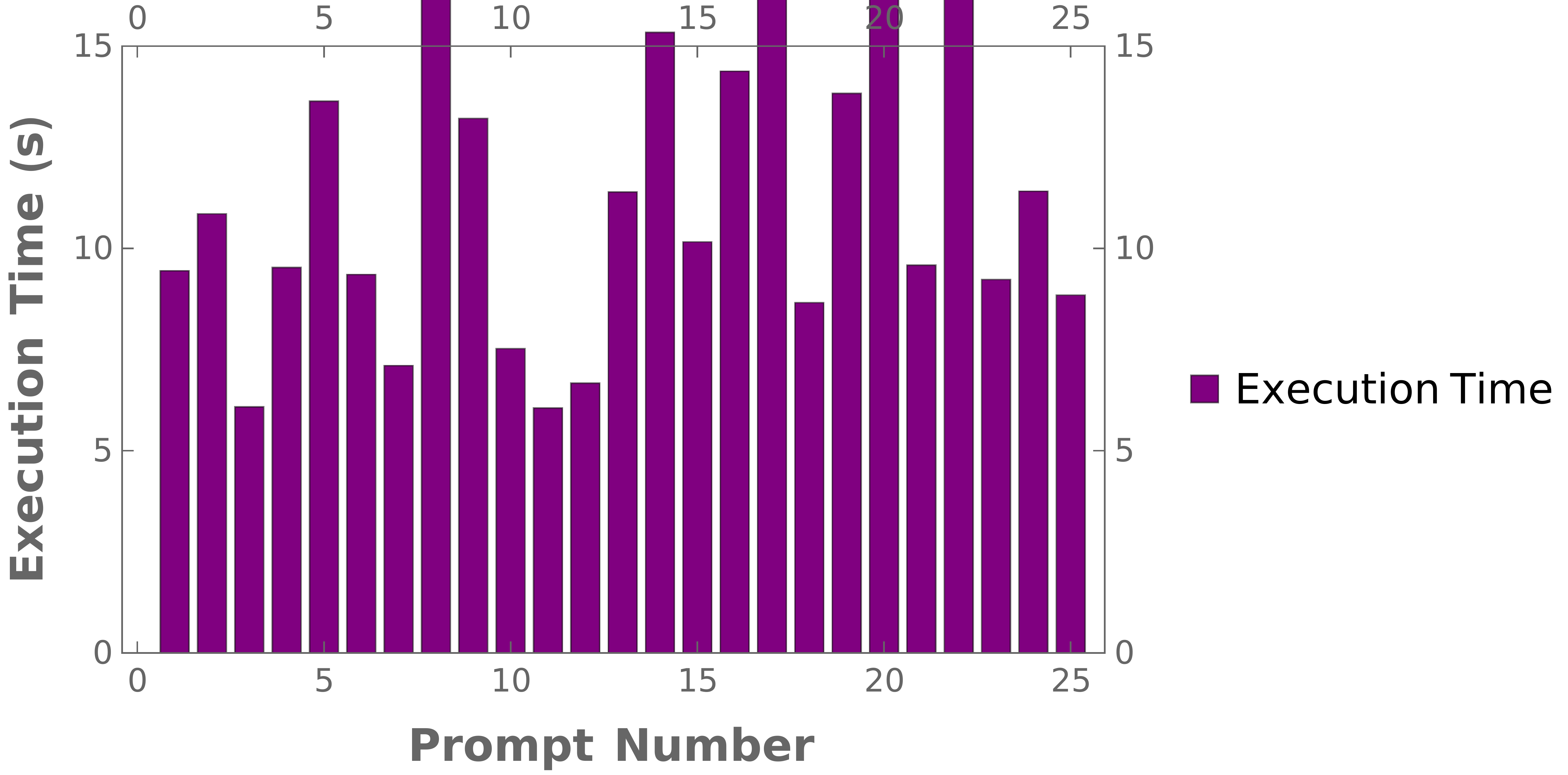}
  \hfill
  \includegraphics[width=0.45\textwidth]{AIExecutionTimes/OpenAIGraphvisualization15.png}
  \caption{Comparison DeepSeek (Top) and OpenAI (Bottom)}
  \label{fig:DeepSeekAIExecutionTimeComparison}
\end{figure}

Gemini was followed by WolframLLM, OpenAI, and DeepSeek, respectively. Because DeepSeek was extraordinarily slower than the other LLMs on the 0-15 second scale, we elected to use a larger 0-35 second window to accommodate this extra latency, as shown in Figure \ref{fig:DeepSeekAIExecutionTime}. To showcase this difference, we also compared using the original 15-second scale with OpenAI; note the difference in Figure \ref{fig:DeepSeekAIExecutionTimeComparison}.

\begin{center}
\textbf{\Large Average LLM Execution Time Table}
\end{center}
\begin{table}[H]
\centering
\begin{tabular}{| p{3cm} | >{\centering\arraybackslash} p{2.5cm} |}
\hline
\textbf{LLM Model} & \textbf{Average Execution Time (s)} \\
\hline
Gemini & 2.67 \\
WolframLLM & 3.48 \\
OpenAI & 5.22 \\
DeepSeek & 12.23 \\
\hline
\end{tabular}
\caption{Average LLM Execution Time Per Model (Rounded to 3 Decimal Places, in seconds)}
\label{tab:AvgExecutionLLM}
\end{table}

\subsection{ServiceConnect Times}

Next, ServiceConnect times were compared. These are composites of the total execution time measured above, but specifically the time that Wolfram Mathematica took to connect to each LLM service. We measured this to give us a picture of the networking latencies that could exist during testing. Because of the volatility in these graphs, we plotted a 5-point moving average of the data, seeing that DeepSeek was the relative average fastest (only just) as measured by number of points (twenty-five points) and their respective times, averaged, as shown in Table \ref{tab:service_times}.

\begin{figure}[H]
    \centering
    \includegraphics[width=1.0\linewidth]{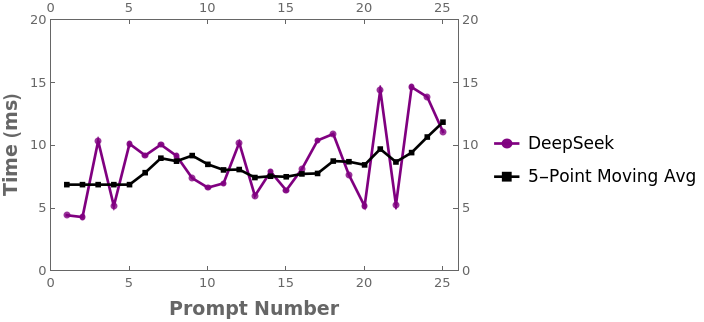}
    \caption{DeepSeek ServiceConnect Moving Averages.}
    \label{fig:OS}
\end{figure}

\begin{figure}[H]
    \centering
    \includegraphics[width=1.0\linewidth]{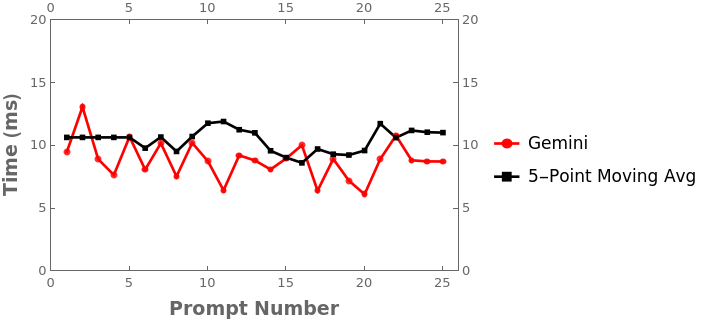}
    \caption{Gemini ServiceConnect Moving Averages.}
    \label{fig:OS}
\end{figure}

\begin{figure}[H]
    \centering
    \includegraphics[width=1.0\linewidth]{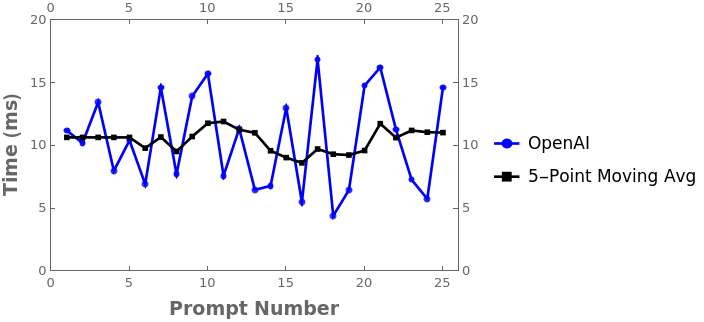}
    \caption{OpenAI ServiceConnect Moving Averages.}
    \label{fig:OS}
\end{figure}

\subsubsection{Three Way Comparison}

\begin{figure}[H]
    \centering
    \includegraphics[width=1.0\linewidth]{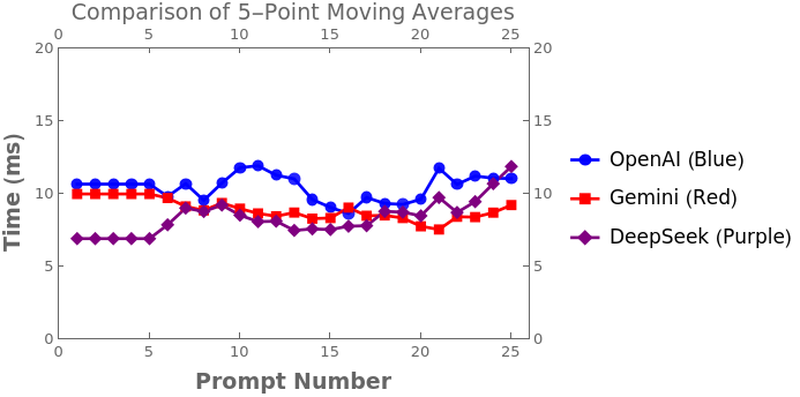}
    \caption{Three Way Three-Point Moving Average Comparison.}
    \label{fig:OS}
\end{figure}

\begin{center}
\textbf{\Large Average Service Connect Time Table}
\end{center}

\begin{table}[H]
\centering
\begin{tabular}{|c|p{3.5cm}|}
\hline
\textbf{LLM Model} & \centering \textbf{Average Service Connect Time (milliseconds)} \tabularnewline
\hline
OpenAI & \centering 10.93 \tabularnewline
Gemini & \centering 8.99 \tabularnewline
DeepSeek & \centering 8.73 \tabularnewline
\hline
\end{tabular}
\caption{Average Service Connect Times for Each LLM (ms)}
\label{tab:service_times}
\end{table}

\subsection{Execution by Python}

To account for any system noise of the testbed, each program was executed twice, with the second execution logged. DeepSeek was first place to execute the most code the fastest, at all prompts under 40 milliseconds.

\begin{figure}[H]
    \centering
    \includegraphics[width=1.0\linewidth]{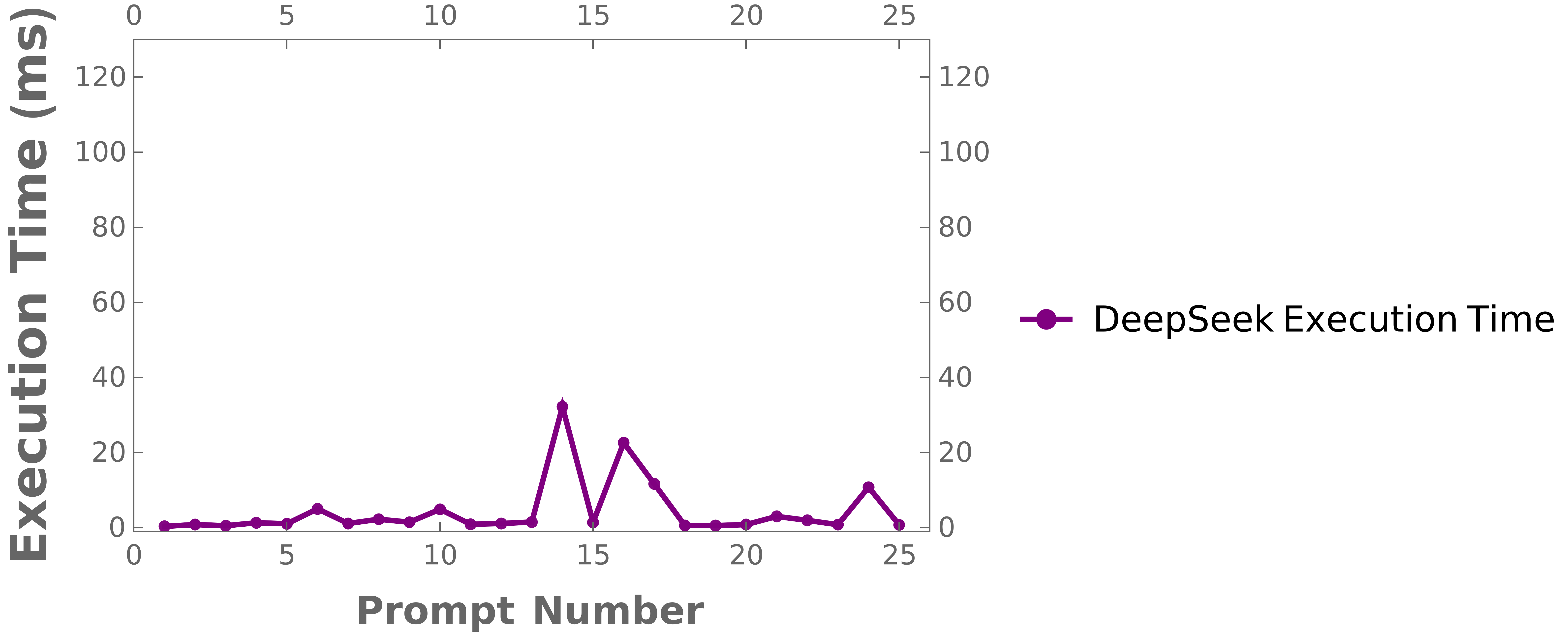}
    \caption{DeepSeek Python Execution Times.}
    \label{fig:OS}
\end{figure}

\begin{figure}[H]
    \centering
    \includegraphics[width=1.0\linewidth]{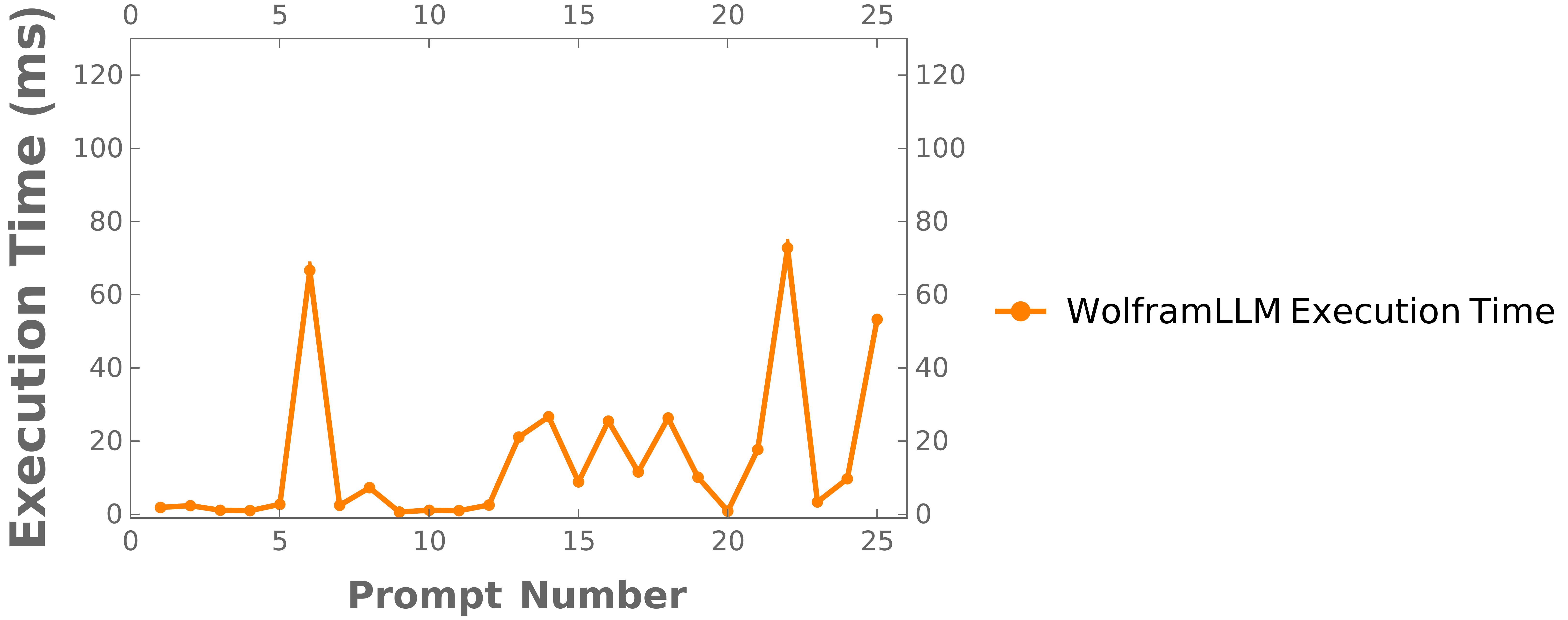}
    \caption{WolframLLM Python Execution Times.}
    \label{fig:OS}
\end{figure}

\begin{figure}[H]
    \centering
    \includegraphics[width=1.0\linewidth]{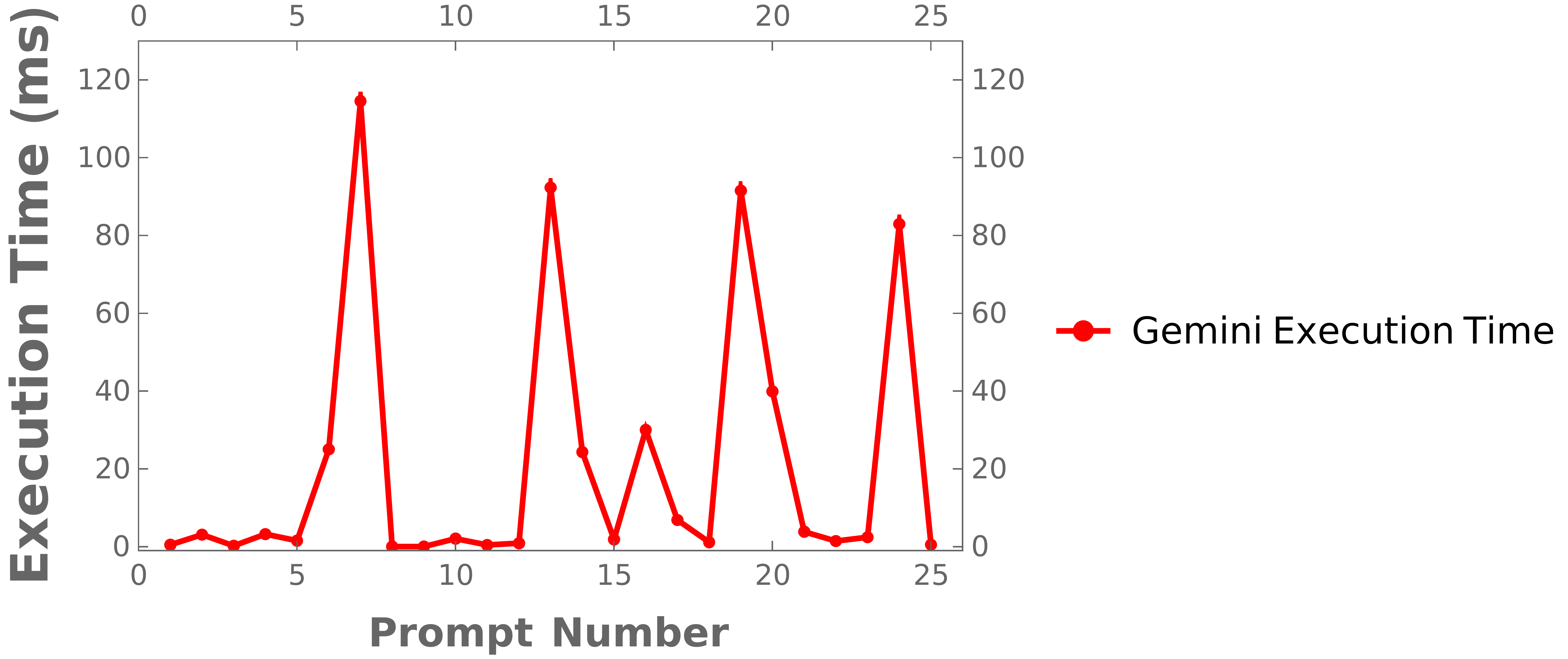}
    \caption{Gemini Python Execution Times.}
    \label{fig:OS}
\end{figure}

\begin{figure}[H]
    \centering
    \includegraphics[width=1.0\linewidth]{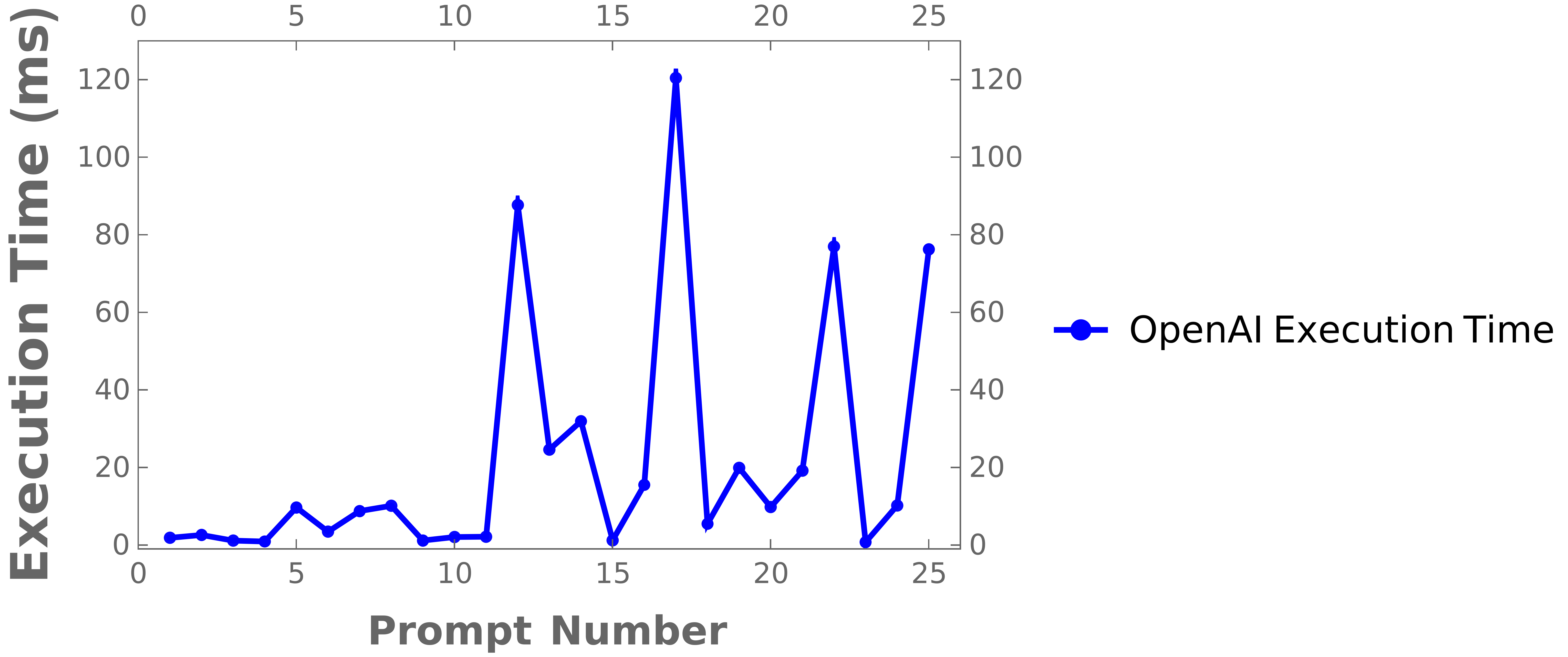}
    \caption{OpenAI Python Execution Times.}
    \label{fig:OS}
\end{figure}

\begin{figure}[H]
    \centering
    \includegraphics[width=1.0\linewidth]{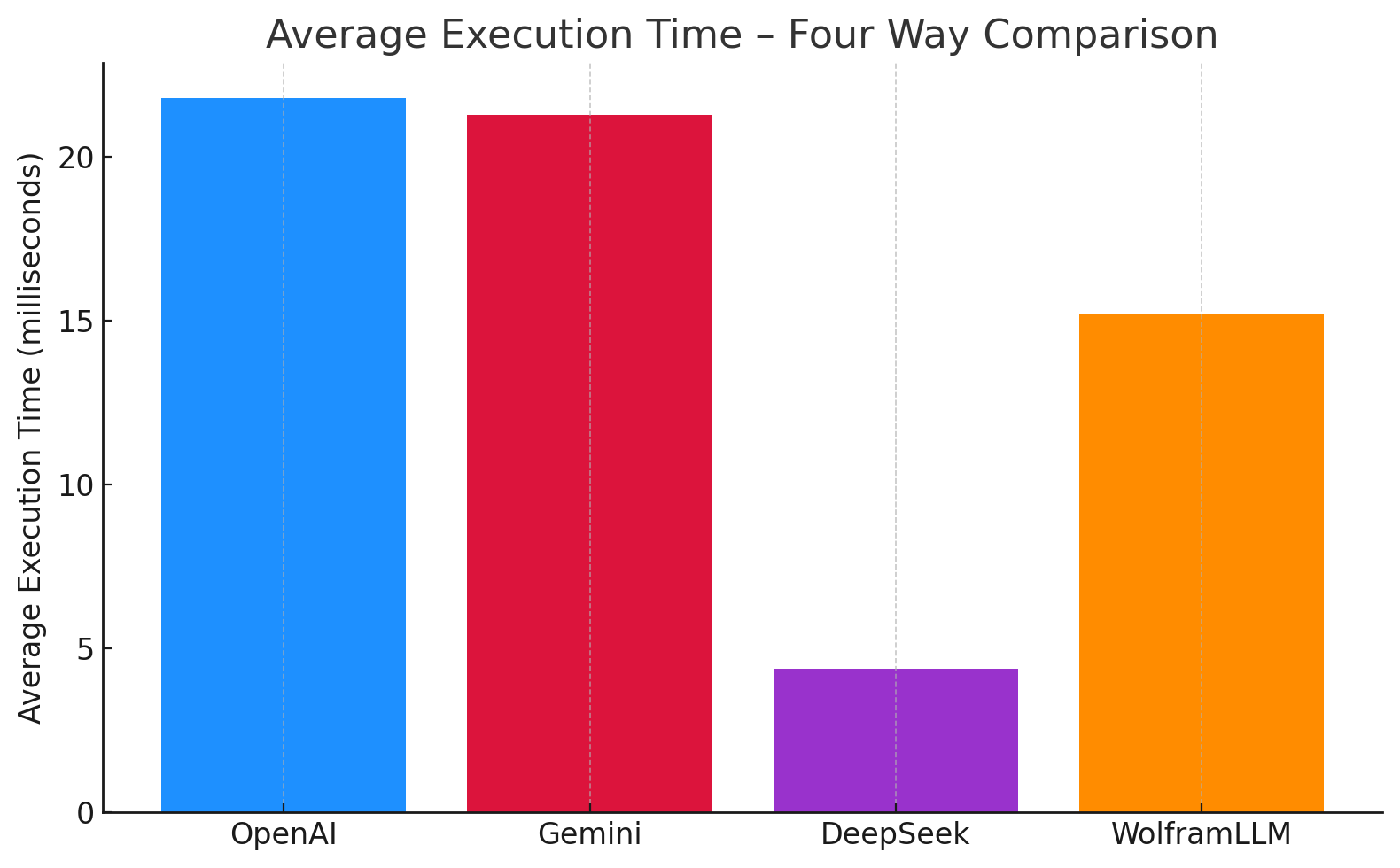}
    \caption{Four Way Average Comparison of Python Execution Times.}
    \label{fig:OS}
\end{figure}

It is noteworthy that of all the LLMs, both OpenAI and Gemini were within a few tenths of a millisecond average of each other across executing all prompts. This average was computed by taking each prompt's execution time, adding it, and divided by the number of prompts (twenty-five). While the averages are very tight, the prompts were of varying execution times (not the same for each prompt).

\begin{center}
\textbf{\Large Python Average Execution Time Table}
\end{center}

\begin{table}[h]
\centering
\begin{tabular}{|l|c|}
\hline
\textbf{LLM Model} & \textbf{Avg. Execution Time (ms)} \\
\hline
DeepSeek & 4.38 \\
WolframLLM & 15.17 \\
Gemini & 21.25 \\
OpenAI & 21.77 \\
\hline
\end{tabular}
\caption{Average Execution Time Per Model (Rounded to 3 Decimal Places, in milliseconds)}
\label{tab:AvgPythonExecutionLLM}
\end{table}

\subsection{Lines of Code (LoC) by Prompt}
Puzzling, both OpenAI and WolframLLM had exactly the same lines of code (averaged) for all programmatic output as shown in Table \ref{tab:avg_loc_models}.

\begin{figure}[H]
    \centering
    \includegraphics[width=1.0\linewidth]{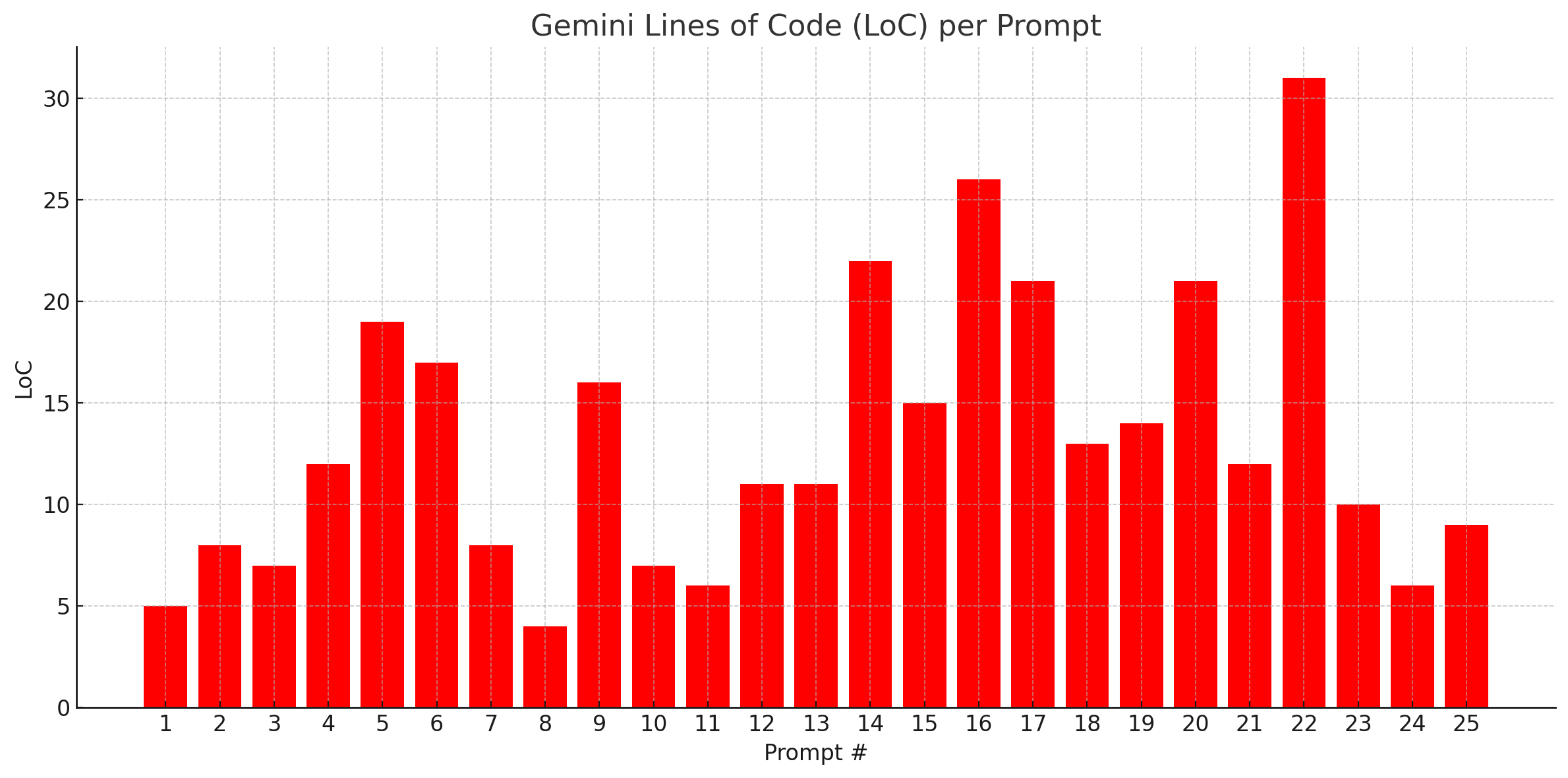}
    \caption{Gemini LoC.}
    \label{fig:OS}
\end{figure}

\begin{figure}[H]
    \centering
    \includegraphics[width=1.0\linewidth]{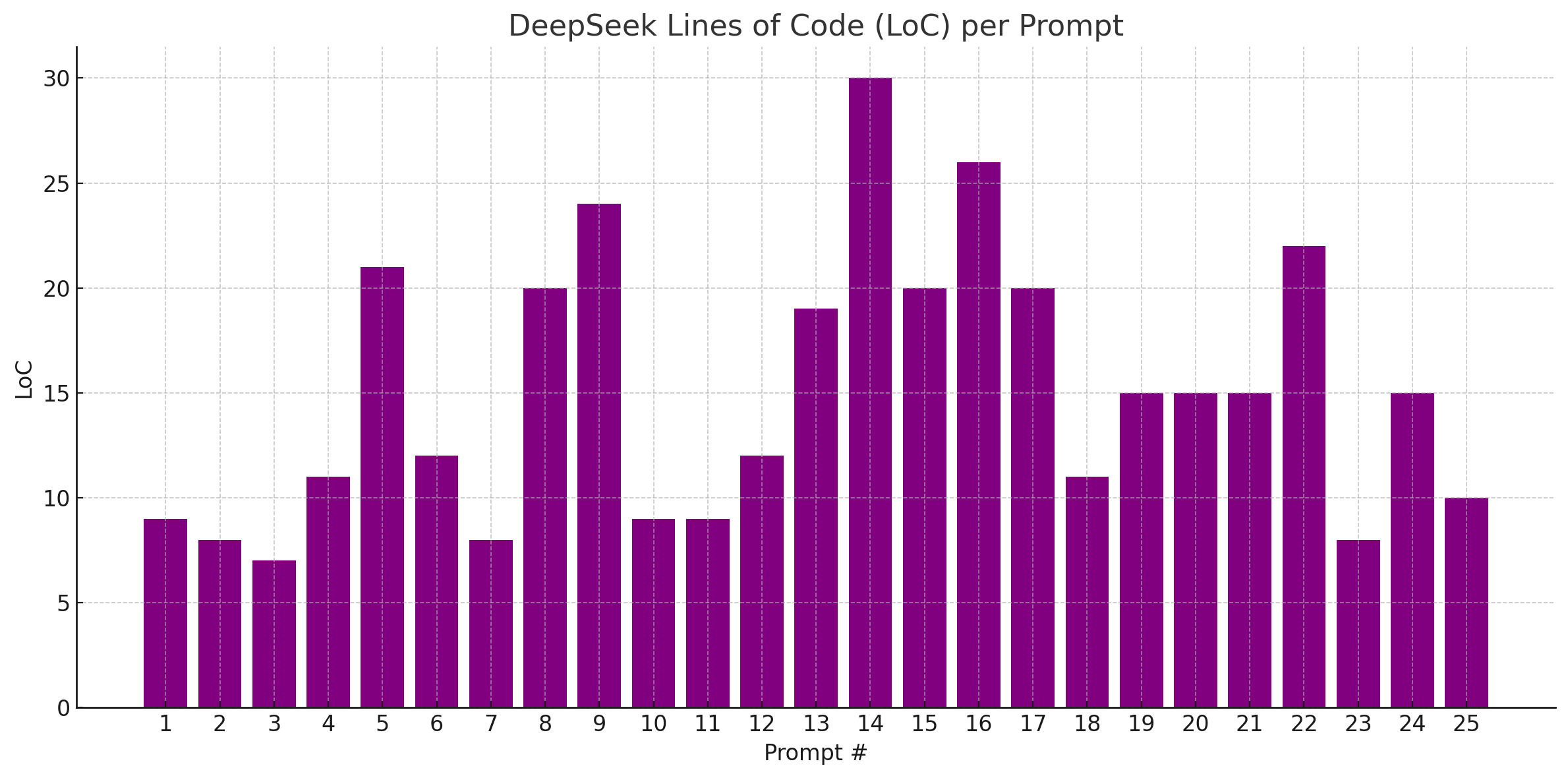}
    \caption{DeepSeek LoC.}
    \label{fig:OS}
\end{figure}

\begin{figure}[H]
    \centering
    \includegraphics[width=1.0\linewidth]{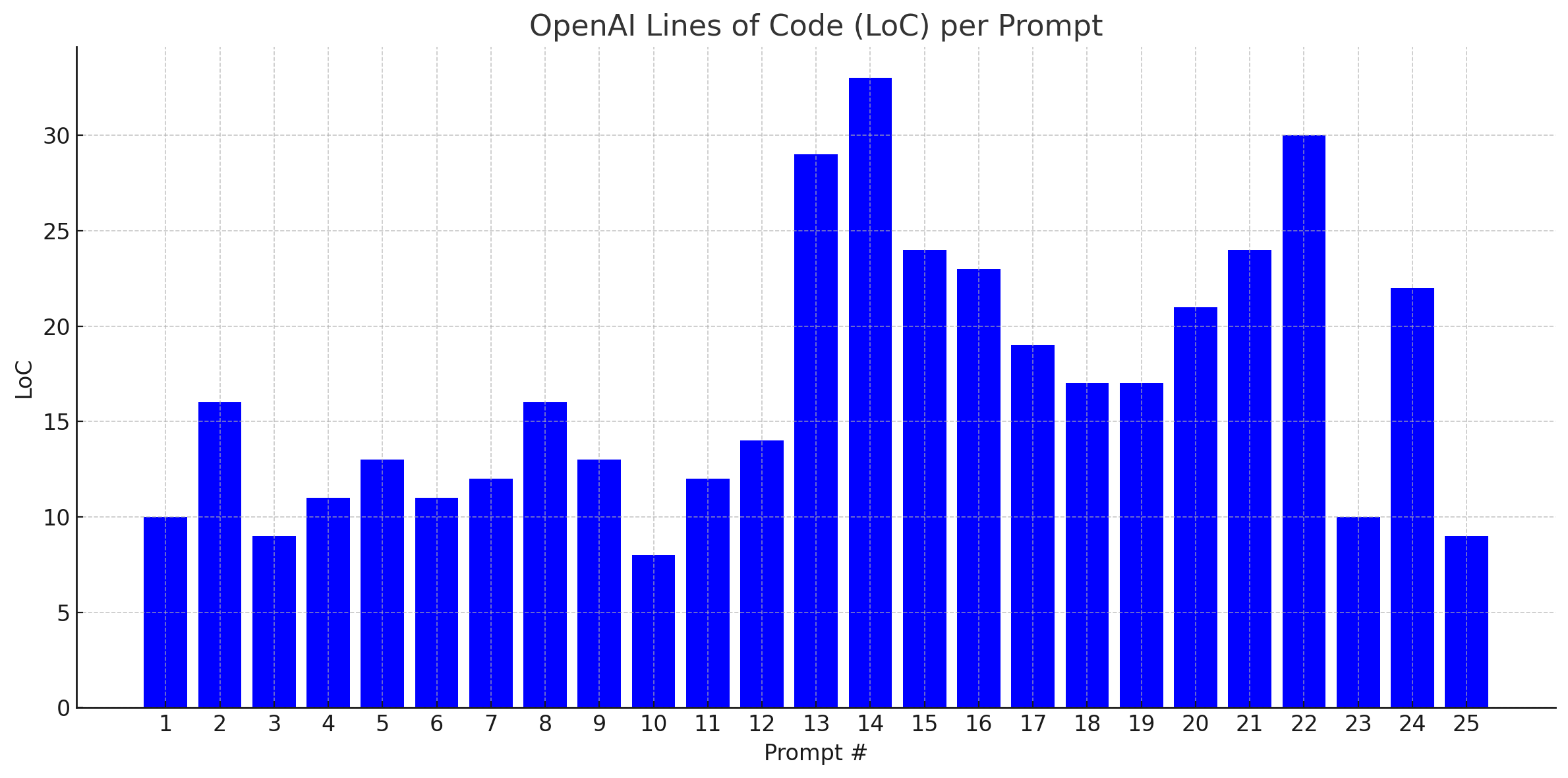}
    \caption{OpenAI LoC.}
    \label{fig:OS}
\end{figure}

\begin{figure}[H]
    \centering
    \includegraphics[width=1.0\linewidth]{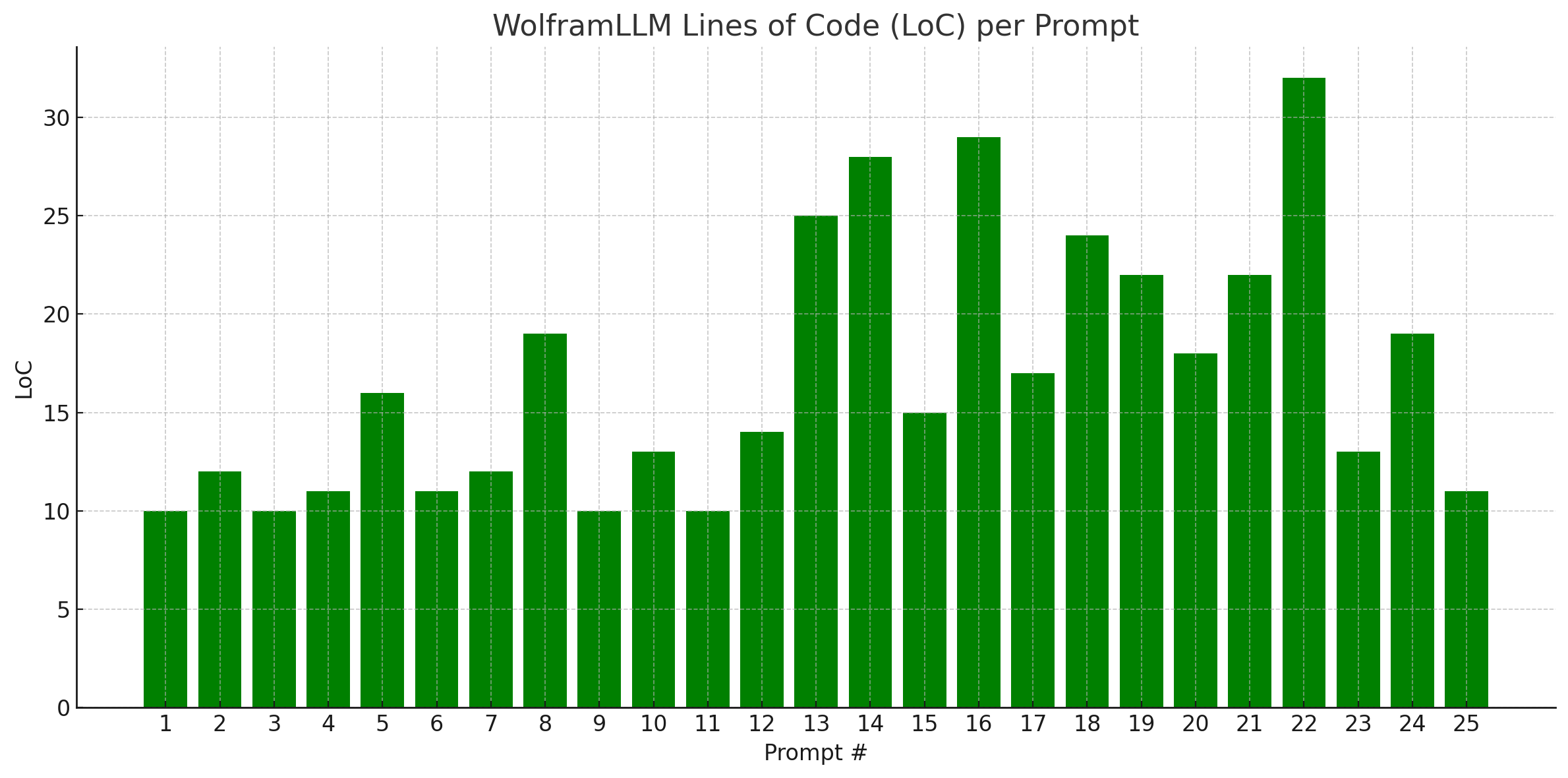}
    \caption{WolframLLM LoC.}
    \label{fig:OS}
\end{figure}

\subsubsection{Average Four Way Comparison}

\begin{figure}[H]
    \centering
    \includegraphics[width=1.0\linewidth]{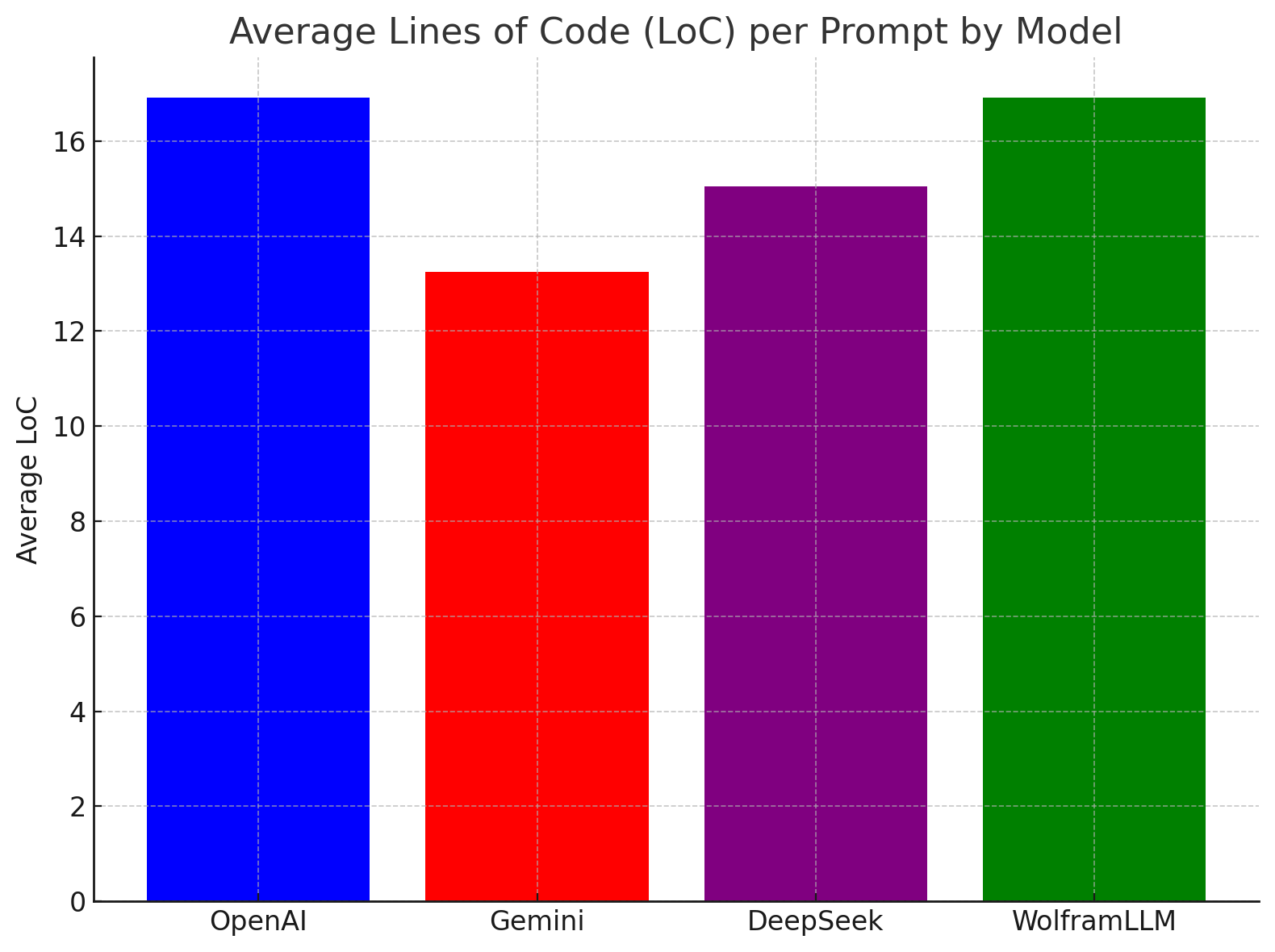}
    \caption{Average Four Way Comparison LoC.}
    \label{fig:OS}
\end{figure}
Important to note, the prompts that failed to compile were different (see Table \ref{tab:errors}), and the size and tallies of each program varied as well. In other words, the outputs were individually different but population-wise, identical!
\begin{center}
\textbf{\Large Average LoC Table}
\end{center}

\begin{table}[H]
\centering
\begin{tabular}{|l|c|}
\hline
\textbf{Model} & \textbf{Average LoC} \\
\hline
Gemini & 13.24 \\
DeepSeek & 15.04 \\
OpenAI & 16.92 \\
WolframLLM & 16.92 \\
\hline
\end{tabular}
\caption{Average Lines of Code (LoC) per Prompt by Model}
\label{tab:avg_loc_models}
\end{table}

\subsection{Execution by Memory Usage}
We also compared memory usage by platform.
\begin{figure}[H]
    \centering
    \includegraphics[width=1.0\linewidth]{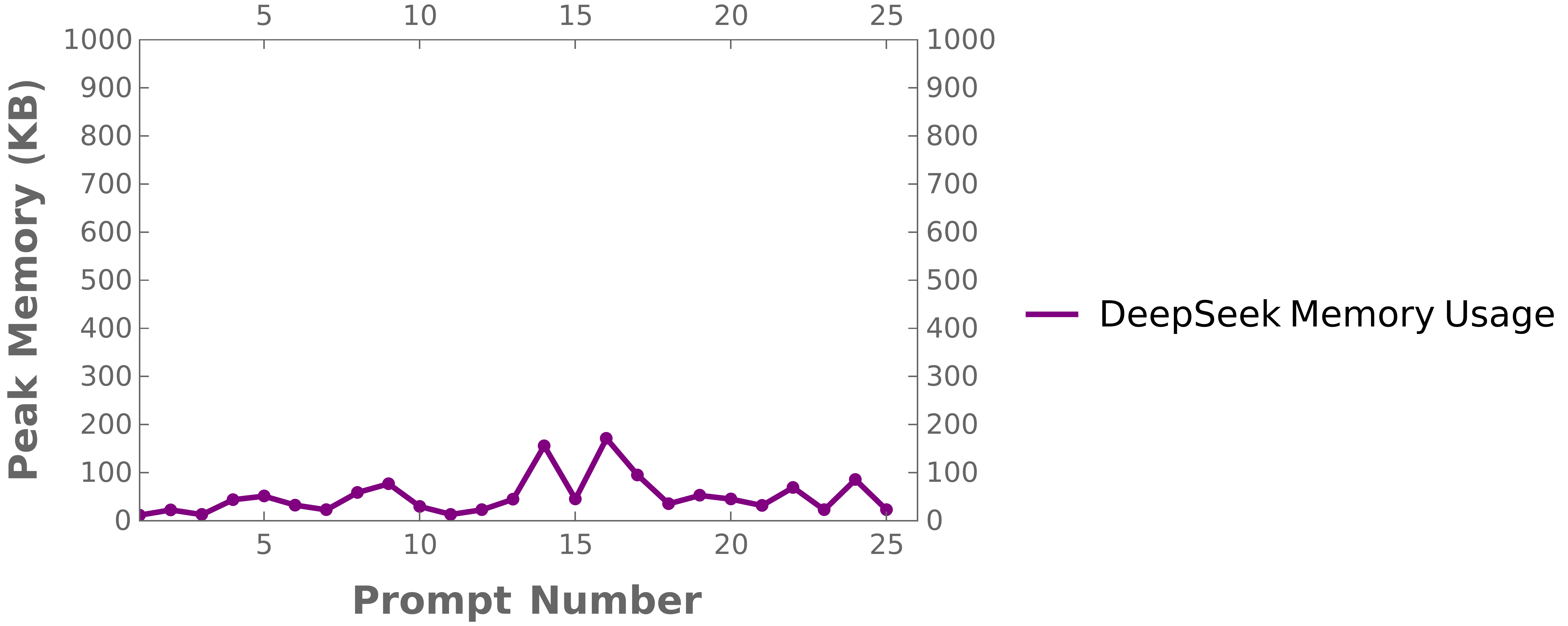}
    \caption{DeepSeek Python Memory Usage.}
    \label{fig:OS}
\end{figure}
\begin{figure}[H]
    \centering
    \includegraphics[width=1.0\linewidth]{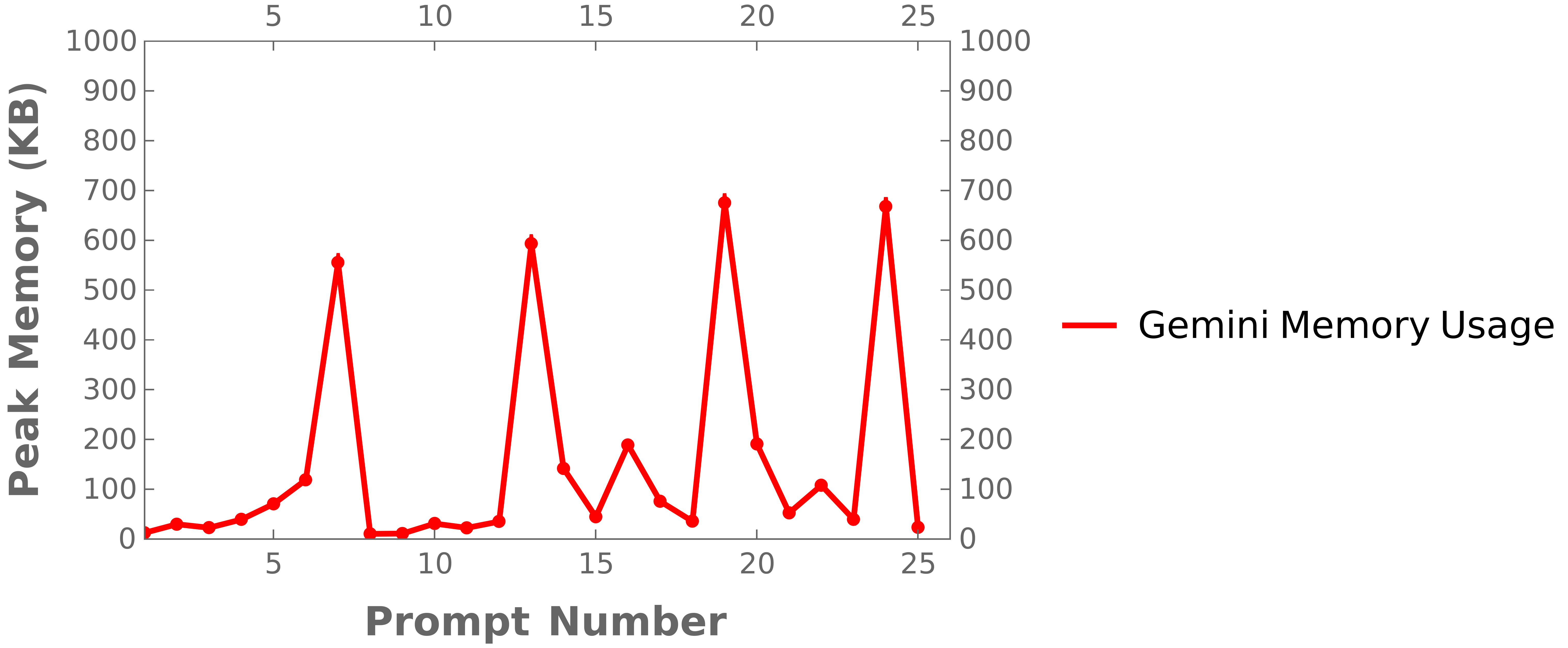}
    \caption{Gemini Python Memory Usage.}
    \label{fig:OS}
\end{figure}
\begin{figure}[H]
    \centering
    \includegraphics[width=1.0\linewidth]{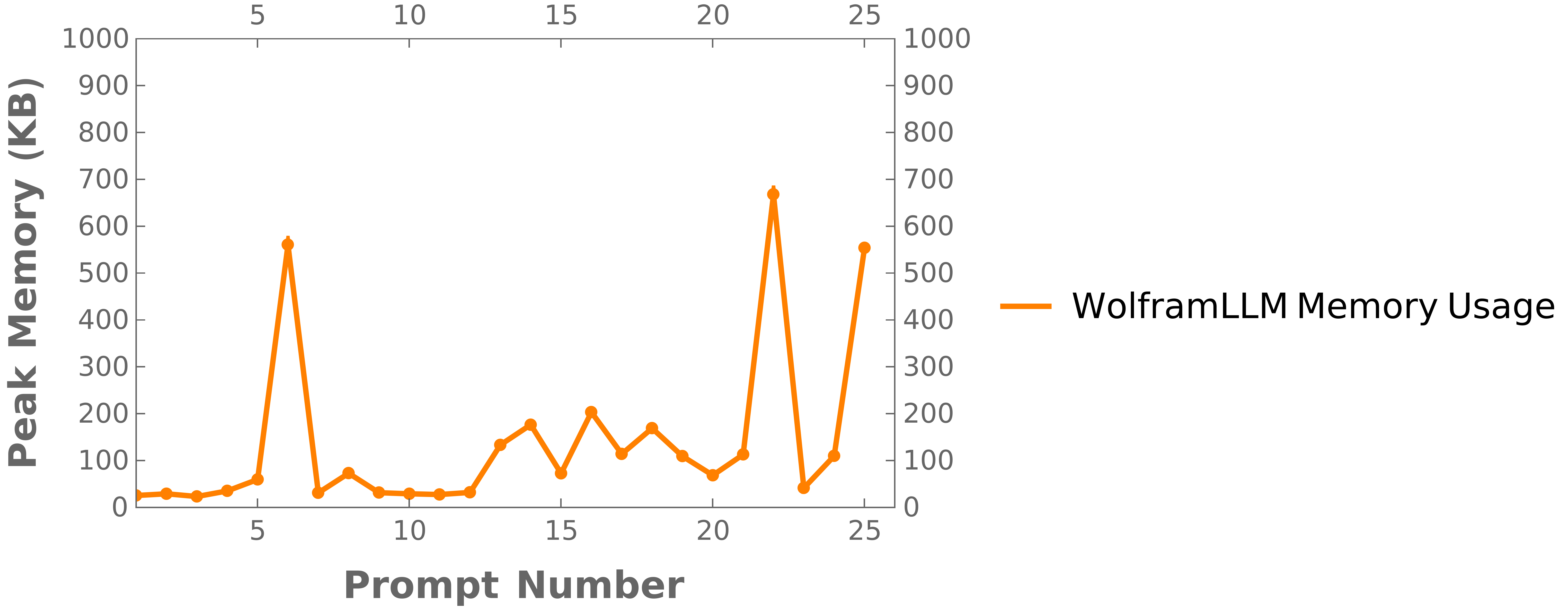}
    \caption{WolframLLM Python Memory Usage.}
    \label{fig:OS}
\end{figure}
\begin{figure}[H]
    \centering
    \includegraphics[width=1.0\linewidth]{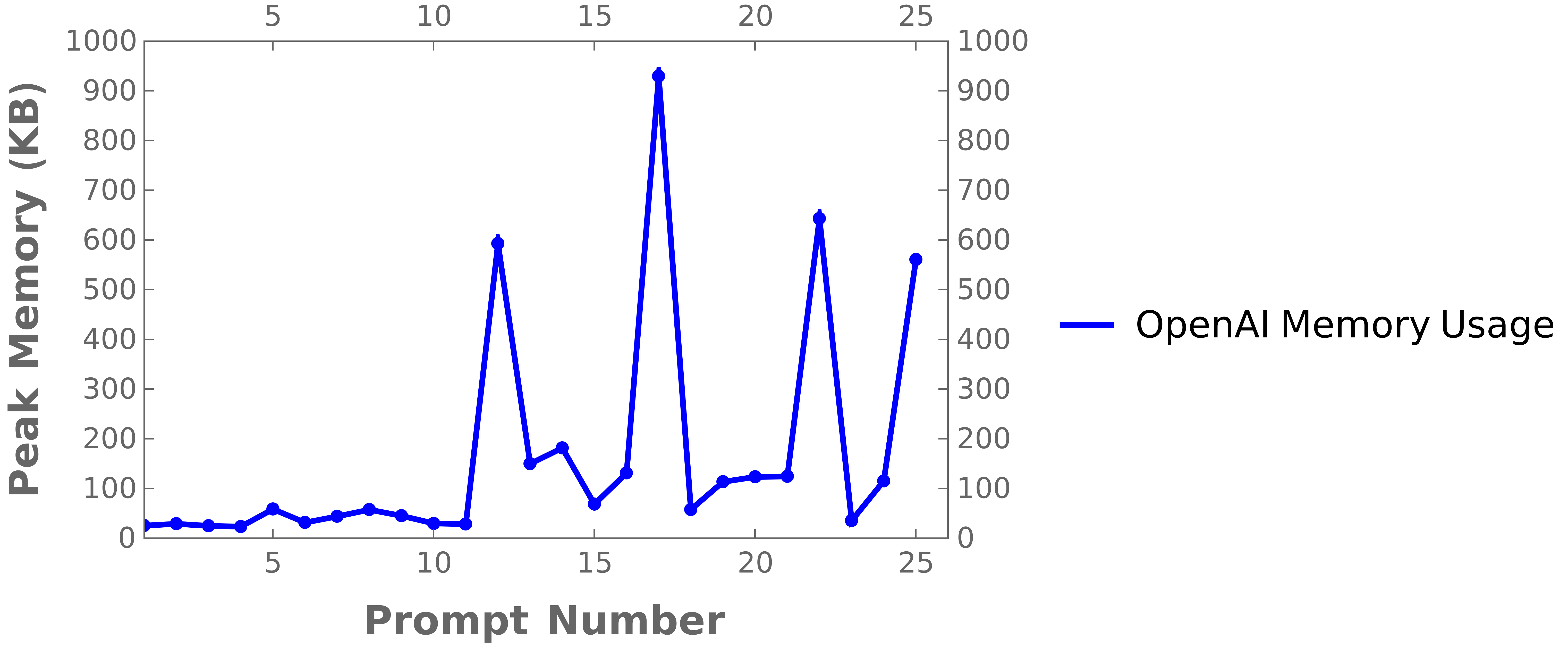}
    \caption{OpenAI Python Memory Usage.}
    \label{fig:OS}
\end{figure}

Following this analysis, we found that the lowest memory usage and resources evaluated according to our testbed was from DeepSeek (as shown in Table \ref{tab:ranked_avg_memory_usage}.
\begin{center}
\textbf{\Large Average Memory Usage by LLM Table}
\end{center}
\begin{table}[H]
\centering
\begin{tabular}{|>{\centering\arraybackslash}p{1cm}| p{2.5cm} |>{\centering\arraybackslash}p{3cm}|}
\hline
\textbf{Rank} & \textbf{LLM} & \textbf{Average Memory Usage (KB)} \\
\hline
1 & DeepSeek & 49.90 \\
2 & Gemini & 154.66 \\
3 & WolframLLM & 156.54 \\
4 & OpenAI & 169.59 \\
\hline
\end{tabular}
\caption{Ranked Average Memory Usage of LLMs (KB)}
\label{tab:ranked_avg_memory_usage}
\end{table}

\subsection{Execution Compilation success rate}

\begin{figure}[H]
    \centering
    \includegraphics[width=1.0\linewidth]{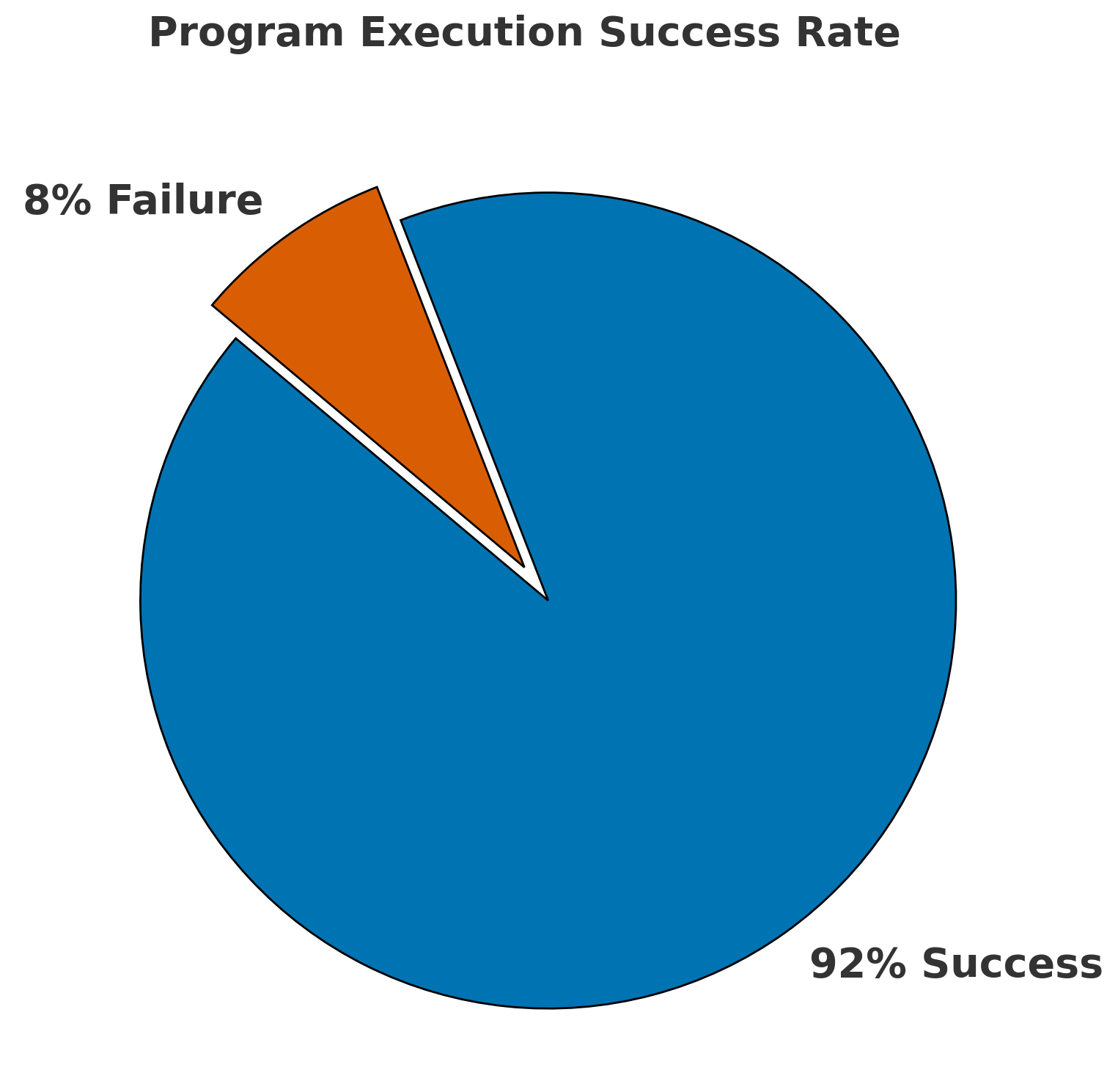}
    \caption{Success rate of all Python prompts for each LLM.}
    \label{fig:LLMSuccessRatePieChart}
\end{figure}

Astonishingly, we found for every LLM, OpenAI, Gemini, WolframLLM and DeepSeek, none performed better overall in their success rate at producing executable code. The best and worst success rate of every LLM was 80\% (as shown in Figure \Ref{fig:LLMSuccessRatePieChart}). This rate stemmed from the fact each LLM had at least two prompts that would not compile.

To curious readers, a list of the prompts that did not compile, with these prompts listed and compared to other LLMs is provided in the appendix. As evident in Table \ref{tab:errors}, no prompt other than Prompt 13 was failed equally across LLMs. Even in this singular case of Prompt 13, the errors between LLMs (DeepSeek V. WolframLLM) are completely different (note a QasmQobj error versus a classical bit index out-of-range error as shown in Table \ref{tab:errors}).

\subsection{Evaluation Of Programmatic output}

The next evaluation was how did the results of the LLMs stack up to expectation? Of the programs compiled, 23 individual programs were compared for each LLM (this is minus two non-compiled programs). Because of how the test harness was devised, evaluation results were a combination of "successes" listed as well as gate designs. The test bed program counted successes at twenty-three of twenty-five for each LLM. Important to note, many of the LLM prompts and outputs relied upon the Qiskit Framework. This discretion was not prompted as evident from the quantum prompt table (available individually in the Appendix as Table \ref{tab:quantum_part1}, Table \ref{tab:quantum_part2}, Table \ref{tab:quantum_part3}, Table \ref{tab:quantum_part4},
Table \ref{tab:quantum_part5}). More about this will be covered in the discussion.

\begin{center}
\textbf{\Large OpenAI Prompt-2 LLM Prompt Output}
\end{center}

\begin{verbatim}
# START
from qiskit import QuantumCircuit
# Create a quantum circuit with 1 qubit
qc = QuantumCircuit(1)
# Apply H gate
qc.h(0)
# Apply Z gate
qc.z(0)
# Apply H gate again
qc.h(0)
# Draw the circuit
print(qc.draw())
# END
\end{verbatim}

\begin{center}
\textbf{\Large OpenAI Prompt-2 LLM In Python Output}
\end{center}

\begin{verbatim}


    q: H Z H

    
\end{verbatim}

Curious readers can see the full programmatic results from each LLM (filtered as well as raw data), including related project files available on a public DukeBox repository: https://duke.box.com/v/pushing-the-limits-of-llms. These are the comprehensive prompts that were generated by OpenAI, Gemini, WolframLLM, and DeepSeek during testing.

\section{Discussion}

Following testing, it became clear from these results that on average, the 4 LLMs, OpenAI, Gemini, WolframLLM, and DeepSeek perform roughly equal to each other in producing working, quantum designs. The meaningful difference is in computation and execution time.

Another important note was that many of the LLM programmatic results relied upon the same Qiskit framework, as opposed to OPENQASM (which Gemini did for Prompt \#9) or manually defining the gates in Python's numpy (which DeepSeek did for prompt \# 4). This suggests the LLM training data may come from the same sources and therefore be bias for quantum operations. A possible cause for this is simply that IBM Qiskit is that much more ubiquitous today in open source literature versus other sources, and henceforth, the training data for LLMs.

In all, if we average the LLM Execution Time data from Table \ref{tab:AvgExecutionLLM} (measured in seconds) and Table \ref{tab:AvgPythonExecutionLLM} (measured in milliseconds), we can see the overall averages in the following Table \ref{tab:ranked_avg_exec_times}: 

\begin{center}
\textbf{\Large Average Overall LLM Table}
\end{center}

\begin{table}[h]
\centering
\begin{tabular}{|>{\centering\arraybackslash}p{1cm}| p{2.5cm} |>{\centering\arraybackslash}p{2.5cm}|}
\hline
\textbf{Rank} & \textbf{LLM} & \textbf{Average Execution Time (s)} \\
\hline
1 & Gemini & 1.35 \\
2 & WolframLLM & 1.75 \\
3 & OpenAI & 2.62 \\
4 & DeepSeek & 6.12 \\
\hline
\end{tabular}
\caption{Ranked Average Execution Times of LLMs (Seconds)}
\label{tab:ranked_avg_exec_times}
\end{table}

This indicates that during testing considering both the LLM's "thinking" execution time, and the actual Python execution time, that Gemini is the fastest overall LLM to produce quantum operations, followed by WolframLLM, and lastly OpenAI. Even if we account for ServiceConnect averages that were available for three of the four LLM models (as shown in Table \ref{tab:service_times}), this still does not change these rankings. In total, the LLMs performed with an accuracy of 80\%. 

These are significant findings, as it demonstrates that current LLMs are both relatively fast, and achieve notable success rates in producing functional quantum operations. But how does this impact quantum gate synthesis in terms of the third key factor introduced earlier-interoperability, alongside reliability and scalability? We anticipate that as LLMs continue to improve, their applications in designing the quantum circuits for quantum computers of the future will expand accordingly. 

Future research should focus on tracking the progression of gate synthesis performance in current and upcoming LLMs, as well as analyzing historical improvements from earlier models. Meanwhile, the data presented here suggest that publicly available LLMs employed for gate synthesis remains an exciting avenue to watch! 
\section{Conclusion}

We presented the first benchmarking study comparing popular and publicly available AI models tasked with creating quantum gate designs. The Wolfram Mathematica framework was used to interface with the 4 LLM AIs, including WolframLLM, OpenAI ChatGPT, Google Gemini, and DeepSeek. This comparison evaluates both the time taken by each AI platform to generate quantum operations (including networking times), as well as the execution time of these operations in Python, within Jupyter Notebook. Our results show that overall, Gemini is the fastest LLM in producing quantum gate designs. At the same time, the LLMs tested achieved working quantum operations 80\% of the time. These findings highlight a promising horizon where publicly available Large Language Models can become fast collaborators with quantum computers, enabling rapid quantum gate synthesis and paving the way for greater interoperability between two remarkable and cutting-edge technologies.



\onecolumn
\appendix

\section{Appendix}

Herein are tables pertaining to the experiments performed.

\begin{center}
\textbf{\Large Prompt Failure And Errors Table}
\end{center}
\begin{table}[H]
\centering
\renewcommand{\arraystretch}{1.3}
\begin{tabular}{|p{2.5cm}|p{4.5cm}|p{5cm}|}
\hline
\textbf{LLM} & \textbf{Prompt File} & \textbf{Error Description} \\
\hline
DeepSeek & Prompt13\_Output.txt & 'Classical bit index 0 is out-of-range.' \\
\hline
DeepSeek & Prompt18\_Output.txt & name 'np' is not defined \\
\hline
Gemini & Prompt9\_Output.txt & invalid syntax (<string>, line 1) \\
\hline
Gemini & Prompt22\_Output.txt & 'Index 1 out of range for size 1.' \\
\hline
OpenAI & Prompt3\_Output.txt & 'InstructionSet' object has no attribute 'num\_...' \\
\hline
OpenAI & Prompt15\_Output.txt & 'duplicate qubit arguments' \\
\hline
WolframLLM & Prompt13\_Output.txt & 'QasmQobj' object has no attribute 'name' \\
\hline
WolframLLM & Prompt20\_Output.txt & 'duplicate qubit arguments' \\
\hline
\end{tabular}
\caption{LLM Prompt Errors and Descriptions}
\label{tab:errors}
\end{table}

\begin{center}
\textbf{\Large Quantum Prompts Table}
\end{center}

\begin{table}[h!]
\centering
\label{tab:quantum_part1}
\renewcommand{\arraystretch}{1.3}
\begin{tabular}{|
>{\raggedright\arraybackslash}p{0.5cm}|
>{\raggedright\arraybackslash}p{2.5cm}|
>{\raggedright\arraybackslash}p{3.5cm}|
>{\raggedright\arraybackslash}p{2.2cm}|
>{\raggedright\arraybackslash}p{2.5cm}|
>{\raggedright\arraybackslash}p{2.0cm}|}
\hline
\textbf{\#} & \textbf{Concept} & \textbf{Prompt} & \textbf{Goal} & \textbf{Expected Output} & \textbf{Eval Criteria} \\
\hline
1 & Identity & Write code that builds a quantum circuit performing the identity operation. The code must begin with \texttt{\# START} and end with \texttt{\# END}. Return only executable code. & No change in state & Empty or canceling gates & Unitary = I \\
\hline
2 & Pauli X & Using only Hadamard and Z gates, write code that implements an X gate. The code must begin with \texttt{\# START} and end with \texttt{\# END}. Output the circuit. & Basis flip & H → Z → H & Gate equivalence \\
\hline
3 & Pauli Y & Write code that constructs a Pauli-Y gate using only X and Z gates. The code must begin with \texttt{\# START} and end with \texttt{\# END}. Output only the quantum code. & Phase + bit flip & iXZ form & Functional correctness \\
\hline
4 & Pauli Z & Write code that expresses the Z gate using a sequence of T and T\textsuperscript{\textdagger} gates. The code must begin with \texttt{\# START} and end with \texttt{\# END}. Return only executable code. & Construct Z from phase gates & T; T; T; T or similar & Phase match \\
\hline
5 & Hadamard & Construct a Hadamard gate from rotation gates using code. The code must begin with \texttt{\# START} and end with \texttt{\# END}. Output only executable code. & X-basis projection & Rx/Rz decomposition & Bloch rotation fidelity \\
\hline
\end{tabular}
\caption{Quantum Prompts Table Part 1 (Prompts 1–5)}
\label{tab:quantum_part1}
\end{table}

\begin{table}[H]
\centering
\label{tab:quantum_part2}
\renewcommand{\arraystretch}{1.3}
\begin{tabular}{|
>{\raggedright\arraybackslash}p{0.5cm}|
>{\raggedright\arraybackslash}p{2.5cm}|
>{\raggedright\arraybackslash}p{3.5cm}|
>{\raggedright\arraybackslash}p{2.2cm}|
>{\raggedright\arraybackslash}p{2.5cm}|
>{\raggedright\arraybackslash}p{2.0cm}|}
\hline
\textbf{\#} & \textbf{Concept} & \textbf{Prompt} & \textbf{Goal} & \textbf{Expected Output} & \textbf{Eval Criteria} \\
\hline
6 & CNOT decomposition & Write code that builds a CNOT gate using only CZ and Hadamard gates. The code must begin with \texttt{\# START} and end with \texttt{\# END}. Return the working circuit. & Functional CNOT & H; CZ; H & Control-target flip \\
\hline
7 & SWAP & Write a quantum circuit that swaps two qubits without using a SWAP gate. The code must begin with \texttt{\# START} and end with \texttt{\# END}. Output only executable code. & Qubit exchange & 3 CNOTs & Measure-ment match \\
\hline
8 & Toffoli (CCNOT) & Decompose the Toffoli gate using only H, T, and CNOT gates. The code must begin with \texttt{\# START} and end with \texttt{\# END}. Write executable code. & Control-control NOT & Standard decomposition & Functional truth table \\
\hline
9 & Fredkin (CSWAP) & Construct a Fredkin (CSWAP) gate using a Toffoli gate or other gates. The code must begin with \texttt{\# START} and end with \texttt{\# END}. Output only the code. & Controlled swap & Toffoli-based or ancilla-based & Swap conditionally \\
\hline
10 & Bell state & Write code that prepares a Bell state between two qubits. The code must begin with \texttt{\# START} and end with \texttt{\# END}. Output the executable circuit only. & Create entanglement & H; CNOT & Result = {00, 11} \\
\hline
\end{tabular}
\caption{Quantum Prompts Table Part 2 (Prompts 6–10)}
\label{tab:quantum_part2}
\end{table}

\begin{table}[H]
\centering
\label{tab:quantum_part3}
\renewcommand{\arraystretch}{1.3}
\begin{tabular}{|
>{\raggedright\arraybackslash}p{0.5cm}|
>{\raggedright\arraybackslash}p{2.5cm}|
>{\raggedright\arraybackslash}p{3.5cm}|
>{\raggedright\arraybackslash}p{2.2cm}|
>{\raggedright\arraybackslash}p{2.5cm}|
>{\raggedright\arraybackslash}p{2.0cm}|}
\hline
\textbf{\#} & \textbf{Concept} & \textbf{Prompt} & \textbf{Goal} & \textbf{Expected Output} & \textbf{Eval Criteria} \\
\hline
11 & Equal superposition & Create a 3-qubit quantum circuit that produces a uniform superposition. The code must begin with \texttt{\# START} and end with \texttt{\# END}. Return only code. & 8-state & $\ket{\psi}$ & 3 Hadamards \\
\hline
12 & GHZ state & Build a GHZ state circuit for 3 qubits using H and CNOT gates. The code must begin with \texttt{\# START} and end with \texttt{\# END}. Output only code. & Multi-qubit entanglement & H + CNOT + CNOT & Result = {000, 111} \\
\hline
13 & W state & Write a circuit that prepares a W state using any method. The code must begin with \texttt{\# START} and end with \texttt{\# END}. Output only the code. & 3-qubit, one-hot entanglement & Controlled superposition & Output = {001, 010, 100} \\
\hline
14 & Deutsch (no CNOT) & Write code that implements Deutsch's algorithm without using the CNOT gate. The code must begin with \texttt{\# START} and end with \texttt{\# END}. Output executable code only. & Functional test of f(x) & H; CZ; H or Oracle variant & 0 = constant, 1 = balanced \\
\hline
15 & Bernstein-Vazirani & Construct the Bernstein--Vazirani algorithm in code for the string s = 101. The code must begin with \texttt{\# START} and end with \texttt{\# END}. Return only the quantum circuit. & Reveal secret string & Hs + Oracle + H & Result = s \\
\hline
\end{tabular}
\caption{Quantum Prompts Table Part 3 (Prompts 11–15)}
\label{tab:quantum_part3}
\end{table}

\begin{table}[H]
\centering
\label{tab:quantum_part4}
\renewcommand{\arraystretch}{1.3}
\begin{tabular}{|
>{\raggedright\arraybackslash}p{0.5cm}|
>{\raggedright\arraybackslash}p{2.5cm}|
>{\raggedright\arraybackslash}p{3.5cm}|
>{\raggedright\arraybackslash}p{2.2cm}|
>{\raggedright\arraybackslash}p{2.5cm}|
>{\raggedright\arraybackslash}p{2.0cm}|}
\hline
\textbf{\#} & \textbf{Concept} & \textbf{Prompt} & \textbf{Goal} & \textbf{Expected Output} & \textbf{Eval Criteria} \\
\hline
16 & Grover's 1-step & Write code for Grover’s algorithm on 2 qubits with a single marked item. The code must begin with \texttt{\# START} and end with \texttt{\# END}. Output executable code only. & Amplify marked amplitude & Oracle + diffuser & Output = target index \\
\hline
17 & Simon’s algorithm & Implement Simon’s algorithm for a 2-bit function in executable code. The code must begin with \texttt{\# START} and end with \texttt{\# END}. Return only the quantum circuit. & Detect hidden XOR pattern & Register logic & Measure-ment structure \\
\hline
18 & Quantum Fourier Transform & Write a 3-qubit quantum Fourier transform circuit in code. The code must begin with \texttt{\# START} and end with \texttt{\# END}. Return only the code. & Phase encoding & Swap + Hadamard + CP gates & Matches QFT matrix \\
\hline
19 & Logical AND (measured) & Construct a quantum circuit that behaves like a logical AND gate via measurement. The code must begin with \texttt{\# START} and end with \texttt{\# END}. Output only the code. & Output = 1 if both inputs = 1 & CCNOT or entangled logic & Measured truth table \\
\hline
20 & Logical OR (measured) & Build a circuit that behaves like a logical OR gate using quantum measurement. The code must begin with \texttt{\# START} and end with \texttt{\# END}. Return executable code. & Output = 1 if any input = 1 & Toffoli + ancilla or phase & Output probability logic \\
\hline
\end{tabular}
\caption{Quantum Prompts Table Part 4 (Prompts 16–20)}
\label{tab:quantum_part4}
\end{table}

\begin{table}[H]
\centering
\label{tab:quantum_part5}
\renewcommand{\arraystretch}{1.3}
\begin{tabular}{|
>{\raggedright\arraybackslash}p{0.5cm}|
>{\raggedright\arraybackslash}p{2.5cm}|
>{\raggedright\arraybackslash}p{3.5cm}|
>{\raggedright\arraybackslash}p{2.2cm}|
>{\raggedright\arraybackslash}p{2.5cm}|
>{\raggedright\arraybackslash}p{2.0cm}|}
\hline
\textbf{\#} & \textbf{Concept} & \textbf{Prompt} & \textbf{Goal} & \textbf{Expected Output} & \textbf{Eval Criteria} \\
\hline
21 & Entangle then swap & Entangle two qubits, then swap them. Write the code that does both. The code must begin with \texttt{\# START} and end with \texttt{\# END}. Output only code. & Confirm state transfer & Bell + SWAP & Measure fidelity \\
\hline
22 & Quantum teleportation & Write a circuit that implements quantum teleportation for a single-qubit state. The code must begin with \texttt{\# START} and end with \texttt{\# END}. Output executable code. & Reconstruct qubit on receiver & Bell + CX + classical correction & Final qubit = input \\
\hline
23 & Controlled-Z from CNOT & Construct a Controlled-Z gate using only CNOT and H gates. The code must begin with \texttt{\# START} and end with \texttt{\# END}. Write the code. & Control-Z logic & H; CNOT; H & Matrix equivalence \\
\hline
24 & XOR by measurement & Write code that encodes XOR into a quantum circuit and decodes by measurement. The code must begin with \texttt{\# START} and end with \texttt{\# END}. Output only the circuit. & 0 if same, 1 if different & CNOT or parity circuit & Output truth table \\
\hline
25 & Random entangled pair & Prepare a random Bell pair from an ancilla and Hadamard gates. The code must begin with \texttt{\# START} and end with \texttt{\# END}. Return only the executable circuit. & Bell state variation & Randomized entangler & Result = {00, 11} \\
\hline
\end{tabular}
\caption{Quantum Prompts Table Part 5 (Prompts 21–25)}
\label{tab:quantum_part5}
\end{table}

\bibliographystyle{IEEEtran.bst}   
\bibliography{bibliography}  


\end{document}